\documentclass[12pt]{iopart}
\usepackage{iopams}  
\usepackage{graphicx}
\usepackage{graphics}
\usepackage{dcolumn}
\usepackage{bm}
\usepackage{subfigure}
\usepackage[utf8]{inputenc}
\usepackage[T1]{fontenc}
\usepackage{mathptmx}
\usepackage[table]{xcolor}
\usepackage{soul}
\usepackage{rotating}

\usepackage[table]{xcolor}
\usepackage{soul}

\usepackage[round,sort&compress]{natbib}

\begin{document}
\bibliographystyle{agsm}
\title[Wake dynamics of parallel foils undergoing abrupt changes in synchronization]{How does altering synchronization of pitching parallel foils change their wake dynamics?}

\author{Ahmet Gungor\textsuperscript{1}, Muhammad Saif Ullah Khalid\textsuperscript{2,3,1}, Arman Hemmati\textsuperscript{1*}}

\address{\textsuperscript{1}Department of Mechanical Engineering, University of Alberta, Edmonton, T6G 1H9, Alberta,          Canada \\
        \textsuperscript{2}Key State Laboratory of Turbulence and Complex Flows, Department of Mechanics, Peking University, Beijing, 100871, People's Republic of China\\
        \textsuperscript{3}Institute of Ocean Research, Peking University, Beijing, 100871, People's Republic of China}

\vspace{10pt}

\begin{indented}
\item{\textsuperscript{*}Corresponding Author: arman.hemmati@ualberta.ca}

\end{indented}
\vspace{10pt}


\begin{abstract}
This study was inspired from the swimming habits of red nose tetra fish that prefer side-by-side configurations and exhibit changes to their synchronization mid-swimming. Using numerical simulations, alterations to the unsteady wake dynamics imposed by abrupt changes in the phase angle between two pitching side-by-side foils were examined at Reynolds number of $4,000$ and Strouhal number of $0.50$. Four hybrid modes were considered in this study with two modes representing an abrupt phase change by $\pi$ during the $20$\textsuperscript{th} cycle. The other two modes represented a simplified case of burst-and-coast swimming, in which there was a brief (2 oscillation periods) suspension of oscillations before imposing a change in phase angle. In all cases, the foils initially performed out-of-phase pitching, and then they started their in-phase motion by either performing the upstroke or down-stroke first. This kinematic change resulted in the formation and growth of a secondary vortex street in between two primary streets, which enabled and maintained a split wake configuration. Furthermore, the phase switching altered the pressure levels on the top and bottom surfaces of both foils to almost similar levels, which attributed to a reduction in the side-force. The growth rate of the secondary vortex street remained consistent for all four hybrid modes.   
\end{abstract}

%
%
\submitto{\BB}
%
%
%

\section{INTRODUCTION}
\label{sec:intro}
Natural species that swim in groups, termed as schools, are ubiquitously found in aquatic environments. Individual members of these specific formations not only gain hydrodynamic benefits, but also they enhance social interactions, expand opportunities to explore better food, and avoid hostile and predating species. Since \cite{weihs-hydromechanics-fish-schooling-1973} and \cite{partridge1979evidence} reported their findings about the collective swimming in different fish species, examining the governing flow mechanisms in fish schools have become the focus of numerous research investigations \citep{herskin1998energy, killen2012aerobic, hemelrijk2015increased, khalid2016hydrodynamics, khalid2018hydrodynamics}. In the context of modern technology, this exploration may also provide effective solutions for various problems related to requirements of energy for robotic swimmers \citep{verma2018efficient, li2019bottom} and tidal \citep{kinsey2012optimal} and wind turbines \citep{kinzel2012energy, brownstein2016performance}. In these studies, researchers have addressed this complex problem related to the hydrodynamics of fish schooling using diverse experimental and numerical techniques based on simplified engineering bodies, such as oscillating foils and panels. In the current study, these insights were used as inspiration  to examine the flow physics associated with the unsteady wake of a simplified side-by-side foil system undergoing an abrupt change in synchronization with applications in autonomous underwater vehicles (AUVs).

Several biologists have proved the energetic advantages attained by schooling members of various aquatic species \citep{fields1990decreased, zuyev1970experimental, herskin1998energy, halsey2018does, killen2012aerobic, ward2018physiology, svendsen2003intra}. Nevertheless, the literature lacks information on experiments for fish-schooling in controlled laboratory environments that should be an essential technique to examine the respective physical mechanisms associated with their kinematics and schooling arrangements \citep{marras2015fish}. For example, \cite{ashraf2016synchronization, ashraf2017simple} performed experiments on the schooling configurations and its benefits for red-nose tetra fish. Swimming with both in-phase and out-of-phase oscillations for two individuals arranging themselves in parallel was observed. They also argued that more synchronized swimming patterns occurred at greater speeds. Under such conditions, this aquatic animal was found to opt the phalanx formation during its swimming \citep{ashraf-simple-phalanx-2017}. Despite many observations in natural swimming of fish, the wake dynamics and vortex interactions that motivate fish to select particular synchronization patterns \citep{burgerhout2013schooling}, or switch from one to the other \citep{ashraf2016synchronization, ashraf2017simple}, remains unexplored. This knowledge is also essential in development and optimal operation of man made systems, such as turbines and AUVs, inspired by natural locomotion of marine species.

Engineers, in conjunction with biologists, have extensively studied simplified models for fish swimming over the years. Using simple oscillatory movements of plates and foils, researchers have revealed different aspects of the governing hydrodynamic principles associated with underwater locomotion of fish, some of which relate to fish schools. In this context, \cite{akhtar2007hydrodynamics} developed a hydrodynamic computational model using two in-line flapping plates, performing both pitching and plunging motion, to analyze the interaction between the dorsal and caudal fins of a bluegill sunfish. The same principles may be extended for two fish swimming in close vicinity of each other. They explained the role of phase relationship between the two foils in order to enhance their thrust production. For a pair of pitching hydrofoils with circular leading-edges, \cite{boschitsch-in-line-tandem-2014} carried out experiments to demonstrate that the gap distance for a pair of pitching foils positioned one behind the other had a significant impact on the propulsive performance of the leading foil. They determined that closely positioned downstream body could cause the augmentation of propulsive performance parameters by around $20\%$. However, the hydrodynamic features were more sensitive to almost all governing kinematic parameters considered. Most importantly, this configuration was more beneficial for the following foil with upto $50\%$ greater thrust and efficiency than those of an isolated foil, if the spacing and phase difference from the leading foil was optimal. An inappropriate selection of these factors might result in $50\%$ performance reduction. Furthermore, several studies \citep{dong2007characteristics, dewey-propulsive-performance-2014, huera2018propulsive, kurt2018unsteady} have reported an enhancement in thrust generation with an improvement in propulsive efficiency when two propellers were arranged in staggered or side-by-side formations with in-phase pitching. With their out-of-phase or anti-phase pitching, hydrodynamic efficiency remained almost the same with greater thrust forces and more power consumption for Reynolds number between $4000$-$11000$. Nevertheless, \cite{dewey-propulsive-performance-2014} also showed that one foil in these formations was able to achieve better thrust and efficiency at the expense of lower hydrodynamic performance for its counterpart. Later on, \cite{bao2017dynamic} determined the dependence of the hydrodynamic characteristics on Strouhal number and spacing between two foils undergoing out-of-phase pitching. They witnessed smaller optimal Strouhal numbers giving a maximized efficiency when the transverse distance was decreased. It was also observed that the transition from drag to thrust producing wakes occurred at smaller Strouhal numbers for shorter spacing distance between foils. In another study, \cite{kurt2018unsteady} presented that propellers pitching synchronously attained a better performance in terms of $20\%$ more thrust and $17\%$ better efficiency with phase differences of $\pi/2$ and $3\pi/2$ when the separation distance was equal to one chord-length. In their study, trends in the variation of power did not coincide with those for thrust, thus leading to a disagreement with the observations by \cite{dewey-propulsive-performance-2014} in this context.

A very interesting aspect of the schooling phenomenon was revealed by \cite{ashraf2016synchronization, ashraf2017simple} during their experiments with red-nose fish. They observed that the members of the school used to tune their tail-fin oscillation frequency to synchronize their swimming either in-phase or out-of-phase in alternate manners. They provided insightful explanations on the effect of swimming configurations and synchronization on the fish propulsive performance. However, the associated wake dynamics mechanisms due to such swimming maneuvers through changes in synchronization of oscillations remain unclear. Furthermore, better understanding of the underlying flow physics associated with the phase and frequency modulation is needed to develop state-of-the-art active control mechanisms to achieve optimal propulsive performance in man-made systems inspired from natural swimming habits of fish.
A recent study by \cite{gungor-2020c} on the change in synchronization of simplified side-by-side pitching foils revealed that varying synchronization patterns enable the foils to maintain their lateral positions inside a multi-foil arrangement. They also reported that the wake configuration after the switch did not follow known patterns for in-phase and out-of-phase pitching foils. Furthermore, it was observed that the foils require a period of $10-20$ cycles to reach quasi-steady performance depending on the on-set of change in synchronization. The current study builds on these findings by exploring the implications of such effects on the unsteady wake dynamics for tandem foils.     

There exists a known correlation between the side (lateral) force production and asymmetric wakes \citep{khalid2015analysis, lagopoulos2020deflected}, for pitching and heaving foils. This may also be relevant to the case of tandem foils undergoing abrupt changes in their pitching synchronization, as also suggested by \cite{gungor-2020c}. In this study, numerical simulations are performed for pitching foils placed in side-by-side configurations inspired by the conditions reported by \cite{ashraf2016synchronization, ashraf2017simple}. The aim of this study is to investigate the changes in wake dynamics and vortex interactions imposed by hybrid-modes of oscillations. This gives a detailed understanding of how structures interact and the wake evolves due to the disturbances imposed by abrupt changes to the synchronization of oscillations in parallel foils. The paper is organized in the following manner. The problem description and computational methodology was provided in section~\ref{sec_problem} followed by the discussion on the results in section~\ref{sec_results}. A brief summary and conclusion of the present study is included in section~\ref{sec_conc}.

\section{COMPUTATIONAL METHODOLOGY}
\label{sec_problem}

\subsection{Geometry \& Kinematics}
The flow around two tandem pitching teardrop hydrofoils in a side-by-side (parallel) configuration was numerically simulated at Reynolds number of 4000, where $Re=U_\infty c/\nu=4000$, $\nu$ is kinematic viscosity, $U_\infty$ is free-stream velocity, and $c$ is the foil chord length. Both foils performed pitching motion with axes passing through their chords at $0.25\mbox{c}$ following:

\begin{equation}
\theta_1(t)=\theta_0\sin(2\pi ft),\label{eq_teta1}
\end{equation}
\begin{equation}
\theta_2(t)=\theta_0\sin(2\pi ft+\phi),\label{eq_teta2}
\end{equation}

\noindent where $\theta_0$ is the pitching amplitude, that is $8^{\circ}$ in the present work, $f$ is frequency of the motion, $t$ is time, and $\phi$ is the phase difference between the two foils. For example, the two pure in-phase and out-of-phase oscillations were simulated by setting $\phi=0$ and $\phi=\pi$, respectively. 

The flow response is commonly characterized using various presentations of the pitching frequency. Strouhal number ($St=fA/U_\infty$, where $A$ is the amplitude of the trailing edge motion) is one of the more common parameters used to identify and characterize the foil oscillations. The parameter space in the current study was inspired by the experiments of \cite{ashraf2016synchronization, ashraf2017simple}, the median of oscillations in which corresponded to $St=0.5$.

\begin{figure}
	\subfigure{\includegraphics[width=1.00\textwidth]{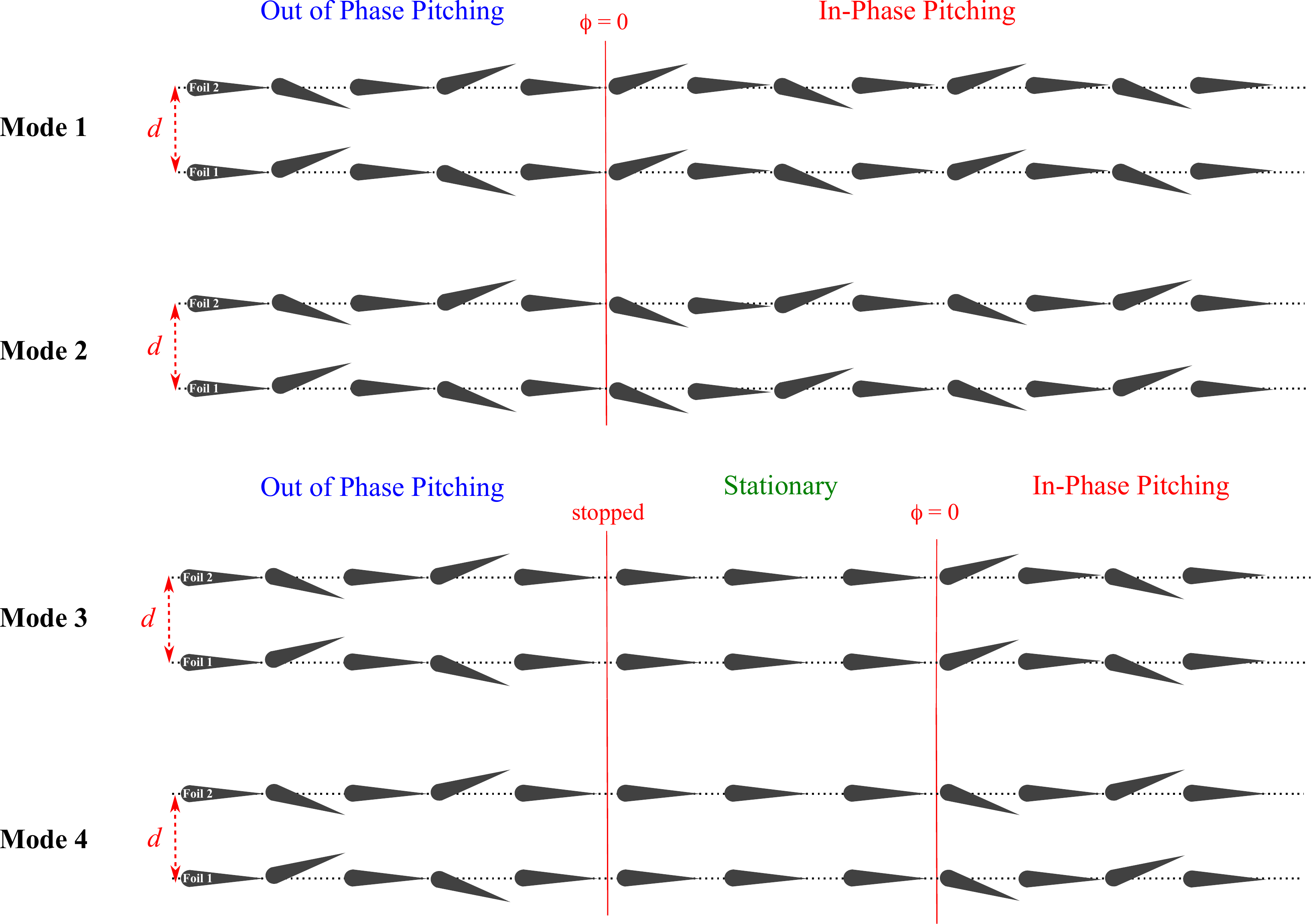}}
	\caption{Graphical demonstration of pitching kinematics for the hybrid oscillation modes}
	\label{fig_motion}
\end{figure}

Four hybrid synchronization modes were considered to identify their implications on both performance and wake dynamics, while cases with pure in-phase and out-of-phase oscillations constituted reference cases. The wake and propulsive performance of each of these cases were investigated for 40 oscillation cycles. Schematics of foils' kinematics for all four modes are provided in Fig.~\ref{fig_motion}. For the out-of-phase pitching, the foils had perfect mirror-image symmetric motion in out-of-phase oscillations, whereas these bodies followed the identical angular path for their in-phase motion. Modes 1$-$4 corresponded to hybrid synchronization modes, in which the foils maintain $\phi=\pi$ for the initial $20$ pitching cycles. In Mode 1 and Mode 2, the change of synchronization (from out-of-phase to in-phase) occurred instantly. The initial direction of the foil's motion after the change of synchronization was clockwise for Mode 1, and the first stroke was in the counter-clockwise direction for Mode 2. The primary objective of this work is to examine and illustrate the underlying flow dynamics, which contributes to how side-by-side foils mitigate wake-deflections and side-force production \citep{gungor-2020c}. The complexity of hybrid modes was further extended, such that Mode 3 and Mode 4 presented cases of delayed swimming with a change in synchronization. Thus, there was a time duration equal to $2\tau$ ($\tau$ is the time-period of pitching oscillations), when oscillations were suspended following the completion of the $20$\textsuperscript{th} pitching cycle. Thereafter, the foils continued their motion by pitching in-phase with a clockwise first stroke for Mode $3$, and counter-clockwise stroke for Mode 4. It is important to note that the foils' kinematics in Mode 3 and Mode 4 were inspired by the burst-and-coast swimming in fish schools \citep{fish-burst-and-coast-1991}. In such cases, the fish swimming is not steady, and the members of schools exhibit intermittent swimming patterns with periods of active oscillation (burst) and gliding (coast). 

\subsection{Numerical Solver}
\label{sec_model}

\begin{figure}
	\subfigure{\includegraphics[width=0.95\textwidth]{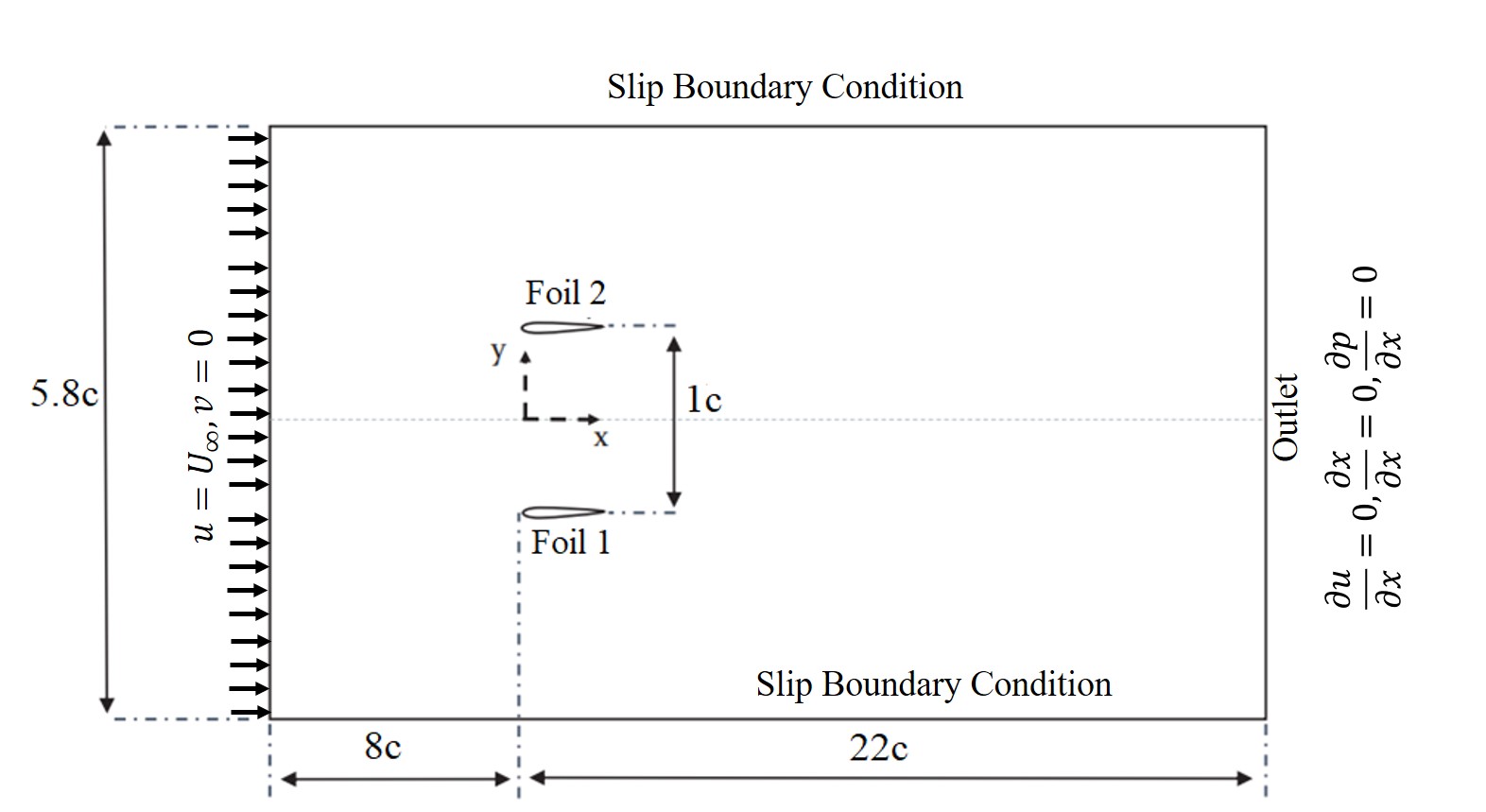}}
	\caption{A schematic representation of the computational domain with boundary conditions (not to scale)}
	\label{fig_domain}
\end{figure}

Flow over two oscillating foils arranged in side-by-side configuration was numerically simulated by directly solving the continuity and Navier-Stokes equations using OpenFOAM. This software package is a well-established finite volume method-based computational fluid dynamics solver that has been validated extensively to simulate underwater swimming problems \citep{csimcsek2020effect, gungor-PRE-2020, hemmati2020reynolds, hemmati2019effect, hemmati2019effects}. Figure~\ref{fig_domain} presents the computational flow domain employed to carry out the present simulations. The rigid teardrop foils had semicircular leading edges with their radii equal to $r=0.05c$ and their separation distance equal to $1c$. The foils were placed at a distance of $8c$ from the inlet of the domain, where the entire domain had a total length of $30c$ in the streamwise direction and a width of $5.8c$ in the cross-stream direction. The boundary conditions were assigned following the simulations of \cite{gungor-PRE-2020}. Uniform velocity ($u=U_\infty,v=0$) was employed as the inlet boundary condition. The outlet boundary condition was set as the Neumann condition for pressure and velocity components in stream-wise and cross-stream directions. The slip boundary condition was used for the upper and lower boundaries, whereas the no-slip boundary conditions were utilized on the surfaces of the two rigid foils. 

PimpleDyMFOAM, which is a transient incompressible flow solver for dynamic meshes, was used to simulate these flows in OpenFOAM. Moreover, the convective and diffusive terms in the Navier-Stokes equations were discretized using a second order central difference scheme. Here, the second-order implicit Backward Euler scheme was employed to numerically approximate temporal terms in the governing equations. The convergence criterion was set as $10^{-5}$, based on the root-mean-square of the residuals of velocity components and pressure in the momentum equation. The maximum Courant number in the domain was bounded to $0.8$ via adequately choosing the time-step size, i.e., there are over $3500$ time-steps per oscillation cycle in these simulations. A non-homogeneous spatial grid was used, which consisted of $~7.21\times10^5$ hexahedral elements. The mesh was concentrated more around the two foils, while the spatial grid expanded toward the boundaries. To handle the unsteady motion of the two foils, the grid was allowed to morph and dynamically adjust itself per time-step according to the formulation for the foils' prescribed pitching kinematics. Further details on the grid-independence studies and validation and verification of the computational methodology can be found in a recent study by \cite{gungor-PRE-2020}. It is important to mention that the verification and validation of the numerical solver was carried out by comparing simulation results with those obtained experimentally by \cite{dewey-propulsive-performance-2014}. The good agreement of the results indicated the effectiveness and accuracy of the presently utilized methodology and solver settings. Moreover, the propulsive performance of the two pitching foils were quantified by various hydrodynamic parameters, including $\widetilde{F}_x$, $\widetilde{M}_z$ and $\widetilde{F}_y$, which were time-averaged over 3500 time-steps for each oscillation cycle \citep{gungor-PRE-2020}. In fish swimming, generation of thrust and side-force as well as the power consumption quantify propulsive performance. Here, cycle-averaged coefficients of thrust ($\widetilde{C}_T$), power ($\widetilde{C}_P$), and side-force ($\widetilde{C}_S$) were used to establish the differences in performance of tandem foils with changing oscillation synchronization. These parameters were defined as:
\begin{equation}
\widetilde{C}_T=\frac{\widetilde{F}_x}{{\textstyle \frac{1}{2}} \rho U_\infty^2 sc},\label{eq_ct}
\end{equation}	
\begin{equation}
\widetilde{C}_P=\frac{\widetilde{M}_z\dot{\theta}}{{\textstyle {1 \over 2}} \rho U_\infty^3 sc},\label{eq_cp}
\end{equation}	
\begin{equation}
\widetilde{C}_S=\frac{\widetilde{F}_y}{{\textstyle \frac{1}{2}} \rho U_\infty^2 sc},\label{eq_cs}
\end{equation}
where $\widetilde{F}_x$ is the streamwise force applied by the foil to the fluid, $\widetilde{M}_z$ is the moment in $z$ direction applied to the foil, $\widetilde{F}_y$ is the crosswise force applied by the foil to the fluid, $\rho$ is fluid density, $s$ is span of the foil. In this study, the time-period of oscillations was defined by $\tau = 1/f$.

\section{RESULTS \& DISCUSSION}
\label{sec_results}
The hydrodynamic performance of the system of side-by-side pitching foils is first studied in detail. This is then expanded into how the wake development for the two foils changes with the change in synchronization between them.   

\subsection{Hydrodynamic Performance}
\begin{figure}
\includegraphics[width=0.95\textwidth]{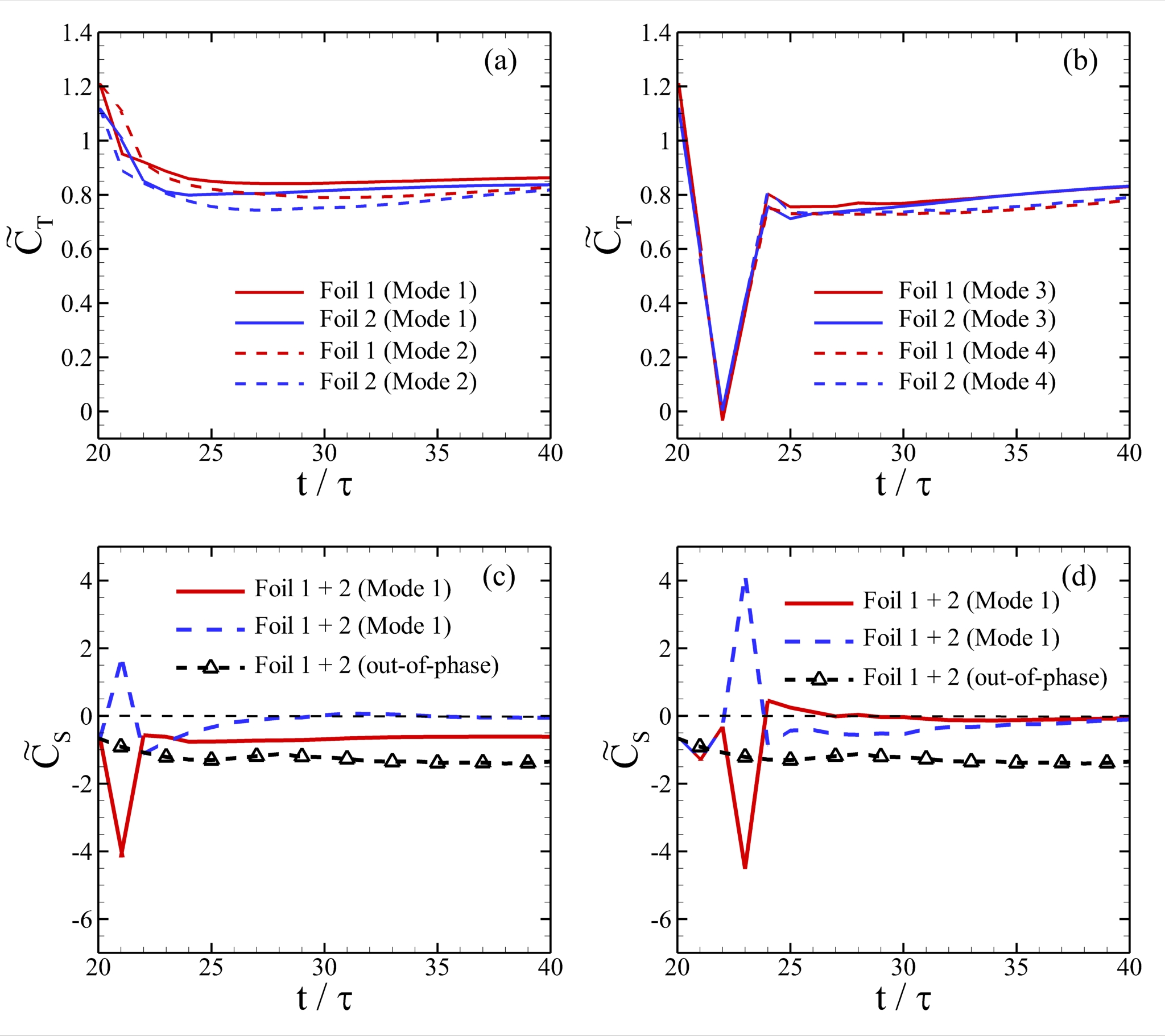}
\caption{Variations of cycle-averaged (a) \& (b) thrust and (c) \& (d) side-force coefficients}
\label{fig_performance}
\end{figure}

The quantification of hydrodynamic performance of the two foils in terms of their cycle-averaged thrust is presented in Fig.~\ref{fig_performance}a for Mode 1 and Mode 2 and Fig.~\ref{fig_performance}b for Mode 3 and Mode 4. Figures~\ref{fig_performance}c and \ref{fig_performance}d exhibited side (lateral) forces for all hybrid modes. Although Gungor and Hemmati \citep{gungor-2020c} has reported these parameters, they are elaborated here as well for completeness. The data for pure pitching cases are also included here for comparison. For all four hybrid modes, both thrust production and power consumption by the foils reduced after switching the out-of-phase motion to in-phase synchronization either instantly (Mode 1 and Mode 2) or after keeping the foils static for some time (Mode 3 and Mode 4). It is evident from Figs.~\ref{fig_performance}a and \ref{fig_performance}b that there was an instant drop of $\widetilde{C}_T$ at the time of switching. Mode 1 and Mode 2 showed more stability towards attaining their steady or quasi-steady states thrust production, respectively. Stationary state of the foils in Mode 3 and Mode 4 for pitching cycles $21$ and $22$ adversely affected their thrust production. However, the system recovered during the $23\textsuperscript{rd}$ cycle after the oscillations were resurrected. Mode $1$ and Mode $2$ produced $25.2\%$ and $27.6\%$ less $\widetilde{C_T}$, respectively, than their counterparts undergoing pure pitching at the end of the 40\textsuperscript{th} cycle. Similarly, Mode 3 and Mode 4 experienced reduction in $\widetilde{C_P}$ by $24.4\%$ and $29.4\%$, respectively. There was also a reduction in $\widetilde{C_P}$ by $29.8\%$, $30\%$, $30.4\%$, and $30.2\%$ for Mode 1 $-$ Mode 4, respectively.   

Next, the unsteady variations of $\widetilde{C}_S$ are shown for all four modes in Figs.~\ref{fig_performance}c and \ref{fig_performance}d. When the foils oscillated purely in out-of-phase, the overall $\widetilde{C}_S$ for the combined dynamical system was negative. It means that the foils experienced pushing force from their upper sides. When these bodies switched their kinematics to in-phase motion after $20$ cycles, the overall $\widetilde{C_S}$ was almost zero for Mode 2, Mode 3, and Mode 4. However, Mode 1 exhibited a significant reduction ($54\%$) in $\widetilde{C_S}$. 

\subsection{Unsteady wake evolution}
Now, the vortex dynamics in the wake of two parallel foils in side-by-side configuration are examined to illustrate how they impact the production of lateral (side-) force. For this purpose, not only the vortex shedding process from the individual foils are described, but also the interference of these vortices, their shape and orientation in the overall wake are explained to establish how they affect the lateral force acting on the foils for all four hybrid modes. 

It is important to note that the pure out-of-phase pitching motion of parallel foils produced two distinct vortex streets formed by pairs of vortices shed from each foil \citep{gungor-PRE-2020}. Initially, these vortex streets traversed in the wake at a larger angle between them. Over time, the angle between the two vortex streets shrank and they started merging into a single vortex street at a distance of $4.5\mbox{c}$ from the foils (see supplementary movie 1). Within the next few oscillation cycles, the onset location of this merging moved upstream and it occurred at $4\mbox{c}$. At this time, two counter-rotating vortices, a vortex with positive vorticity (positive vortex) from the lower array and one with negative vorticity (negative vortex) from the upper array, were trapped by the two vortex streets. Counter-clockwise rotation of the positive vortex and clockwise rotation of the negative vortex induced 
upwash and downwash flow for the lower and upper streets, respectively. Consequently, the wake was deflected upward. After a few more oscillation cycles, the foils' pitching synchronization was abruptly changed, in which case the phase angle between them was switched from $\pi$ (out-of-phase) to $0$ (in-phase). This coincided with a major change in wake behavior around the symmetry line ($y^{+}=0$). The flow development that proceeded with the change of phase angle is explored for each mode in the following subsections.

\begin{figure}
\centering
{\subfigure{\includegraphics[width=0.35\textwidth]{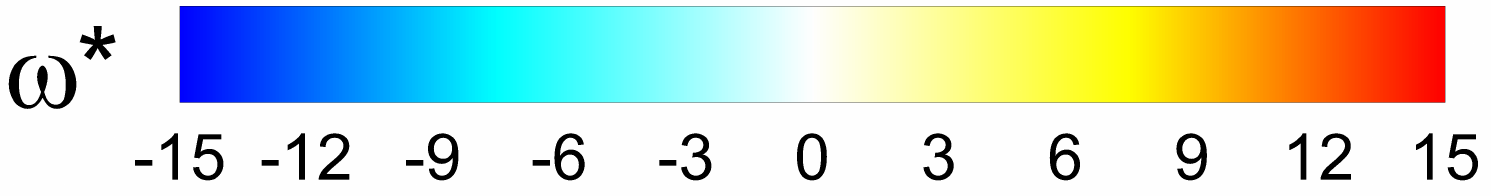}}\\
\subfigure[$\quad t_0=20.5T$]{\includegraphics[width=0.47\textwidth]{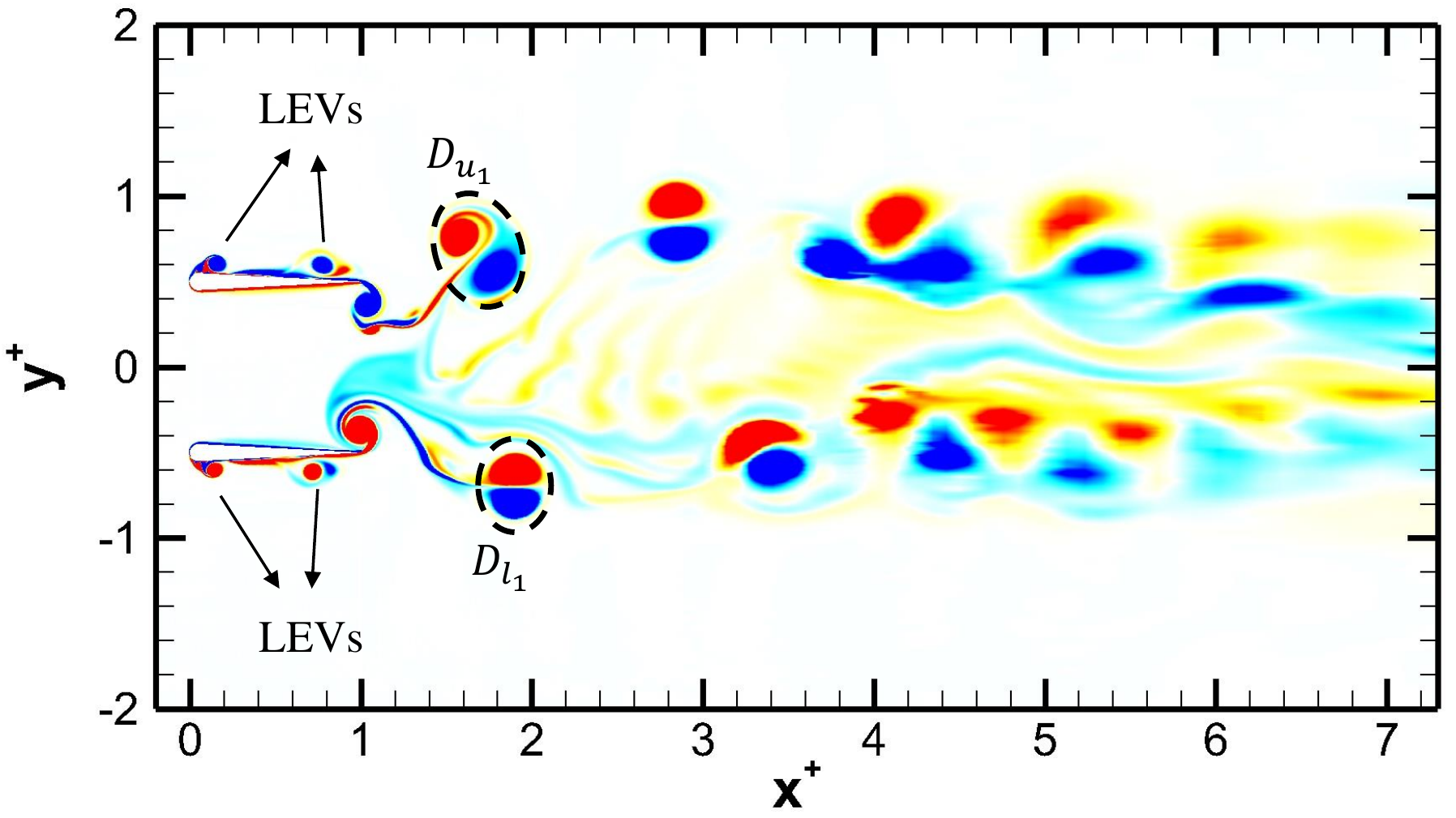}}
\subfigure[$\quad t_1=21T$]{\includegraphics[width=0.47\textwidth]{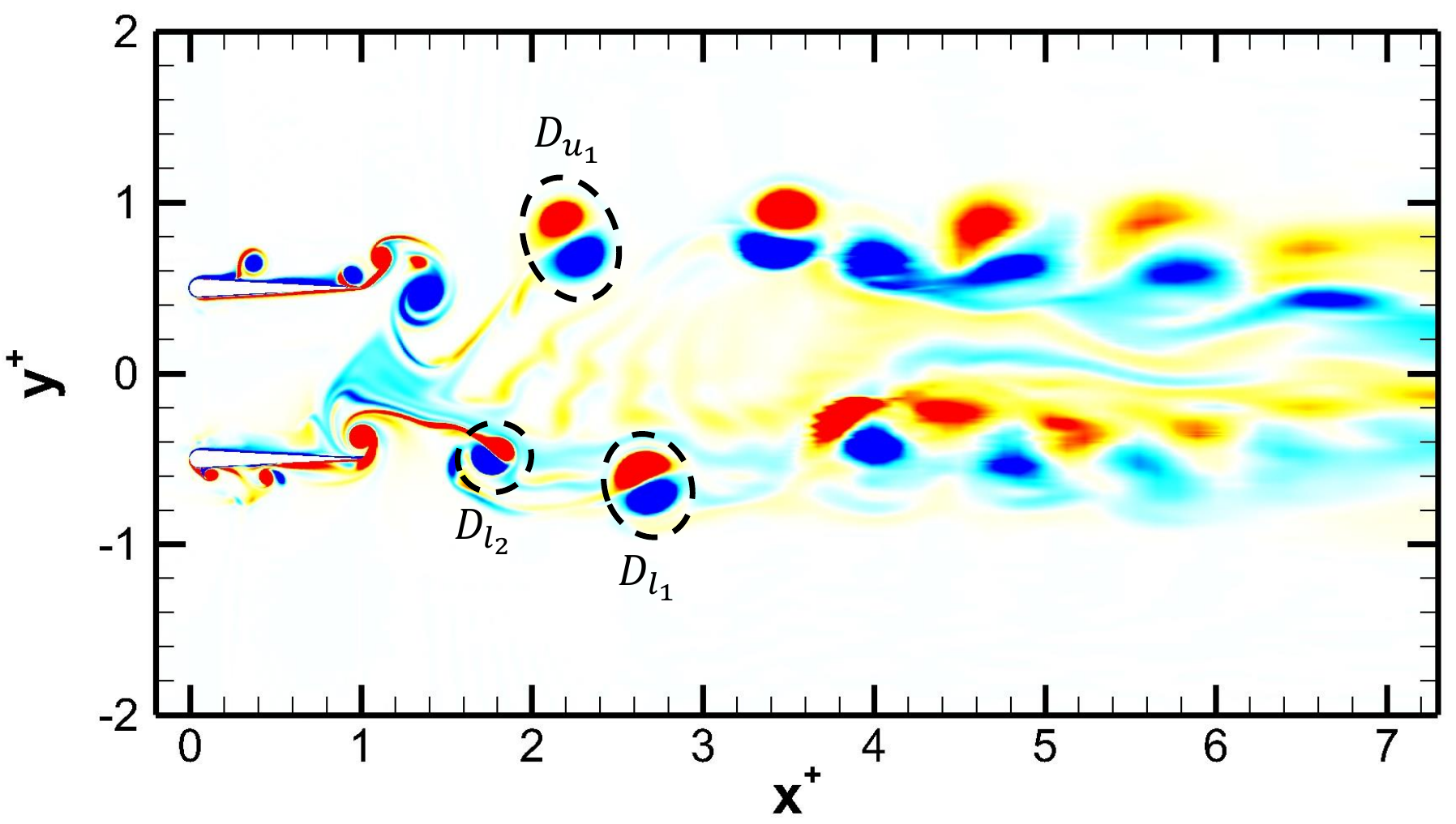}}
\subfigure[$\quad t_2=22T$]{\includegraphics[width=0.47\textwidth]{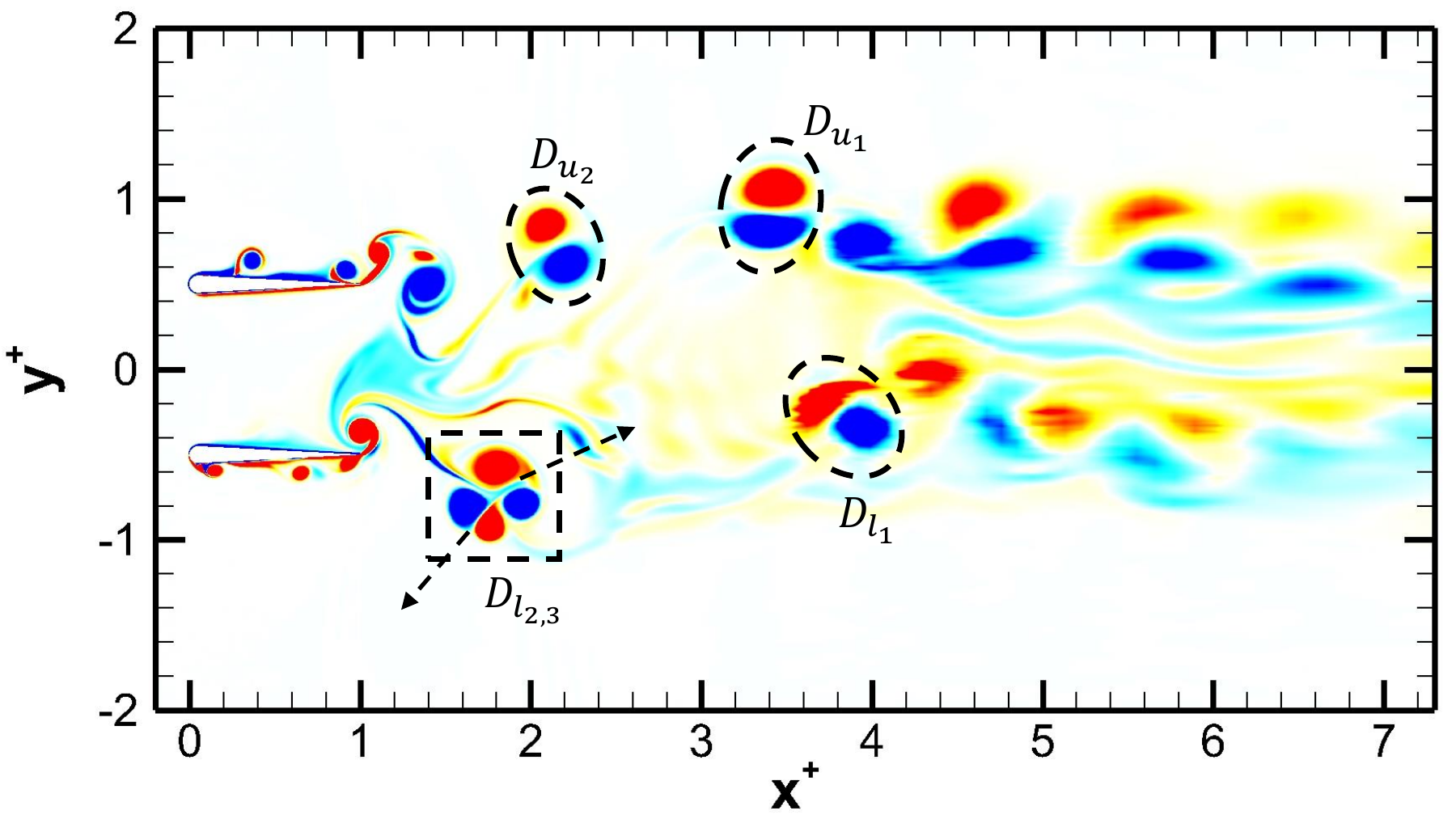}}	
\subfigure[$\quad t_3=23T$]{\includegraphics[width=0.47\textwidth]{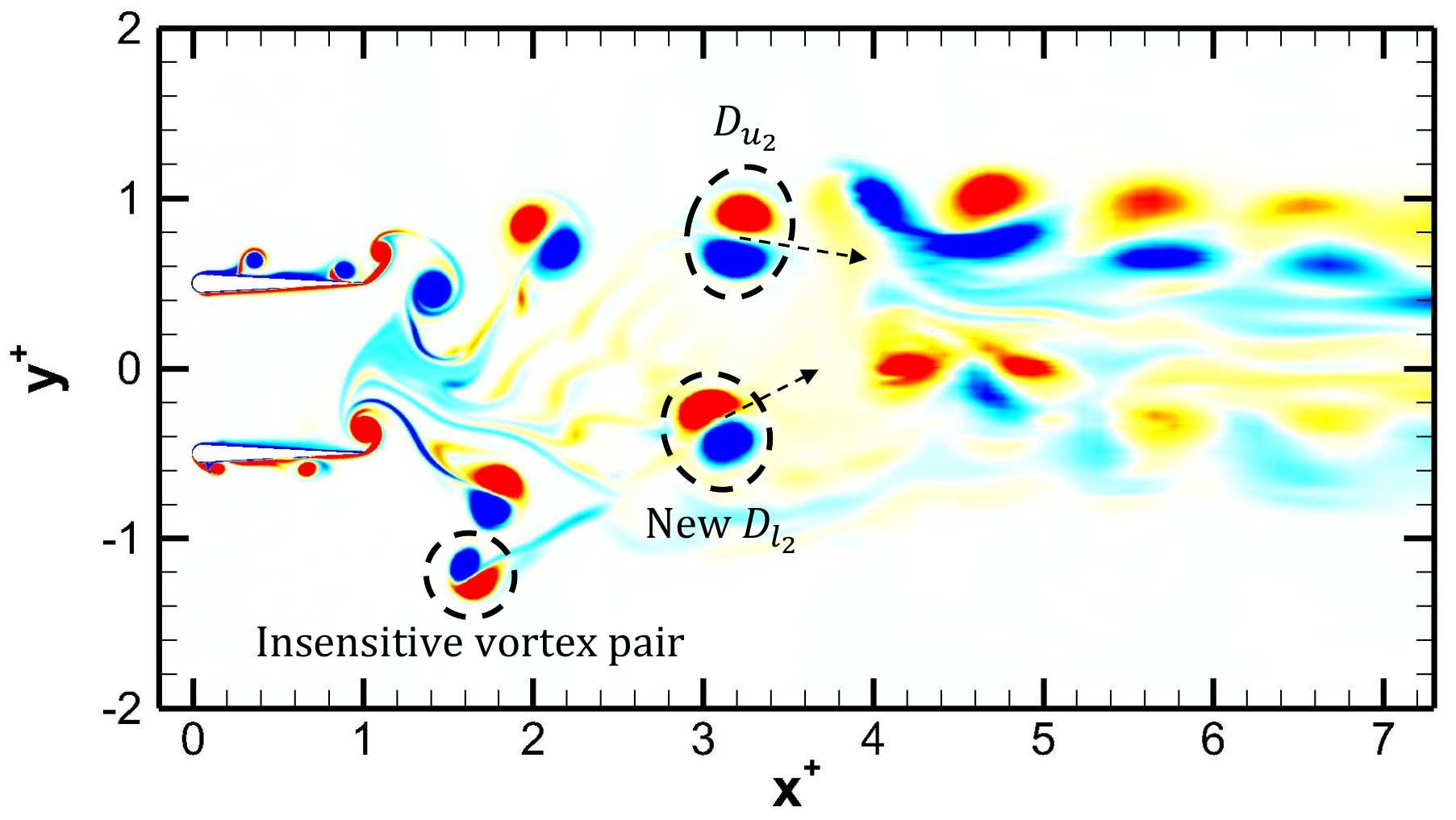}}
\subfigure[$\quad t_4=24T$]{\includegraphics[width=0.47\textwidth]{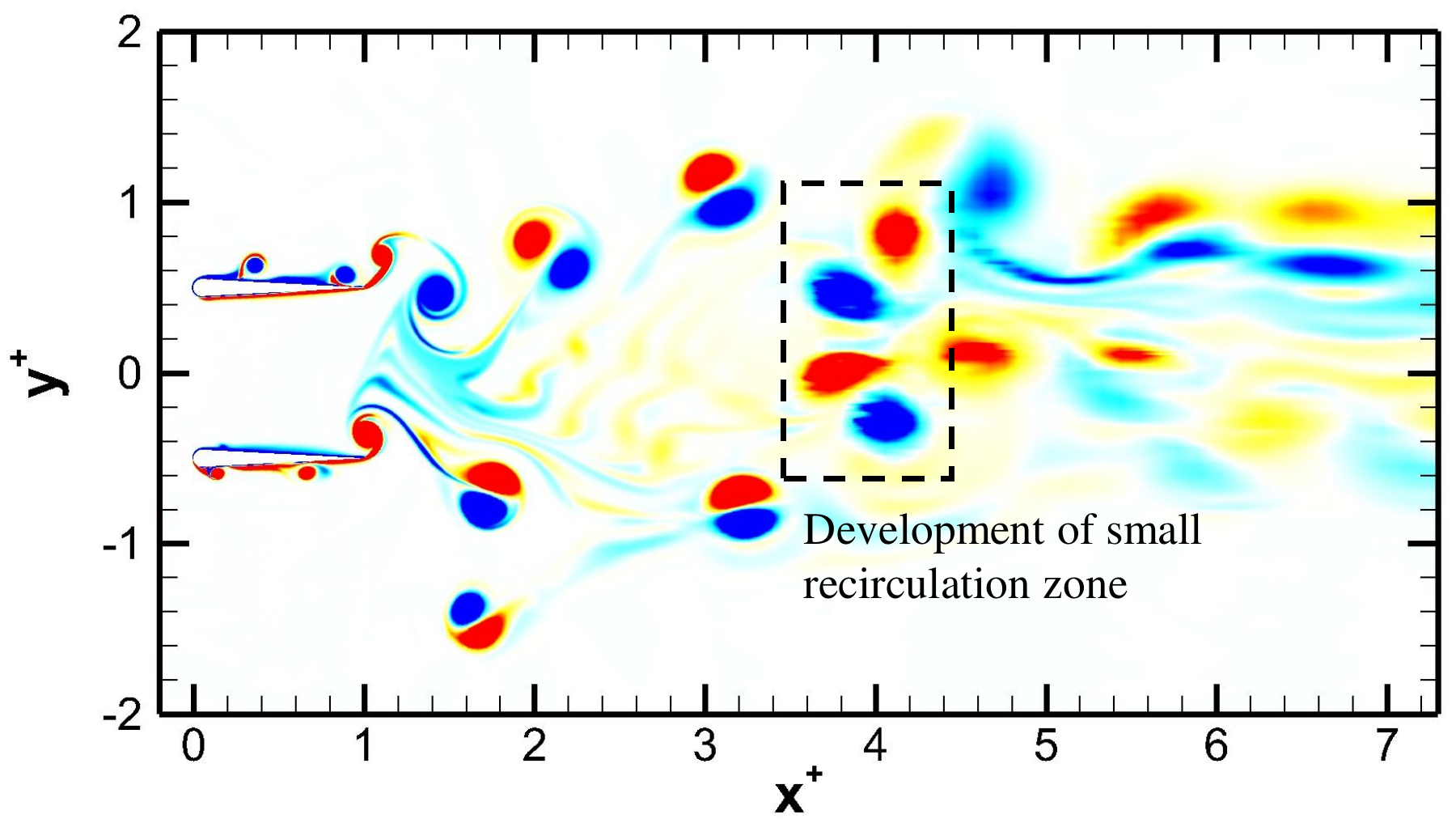}}	
\subfigure[$\quad t_5=27T$]{\includegraphics[width=0.47\textwidth]{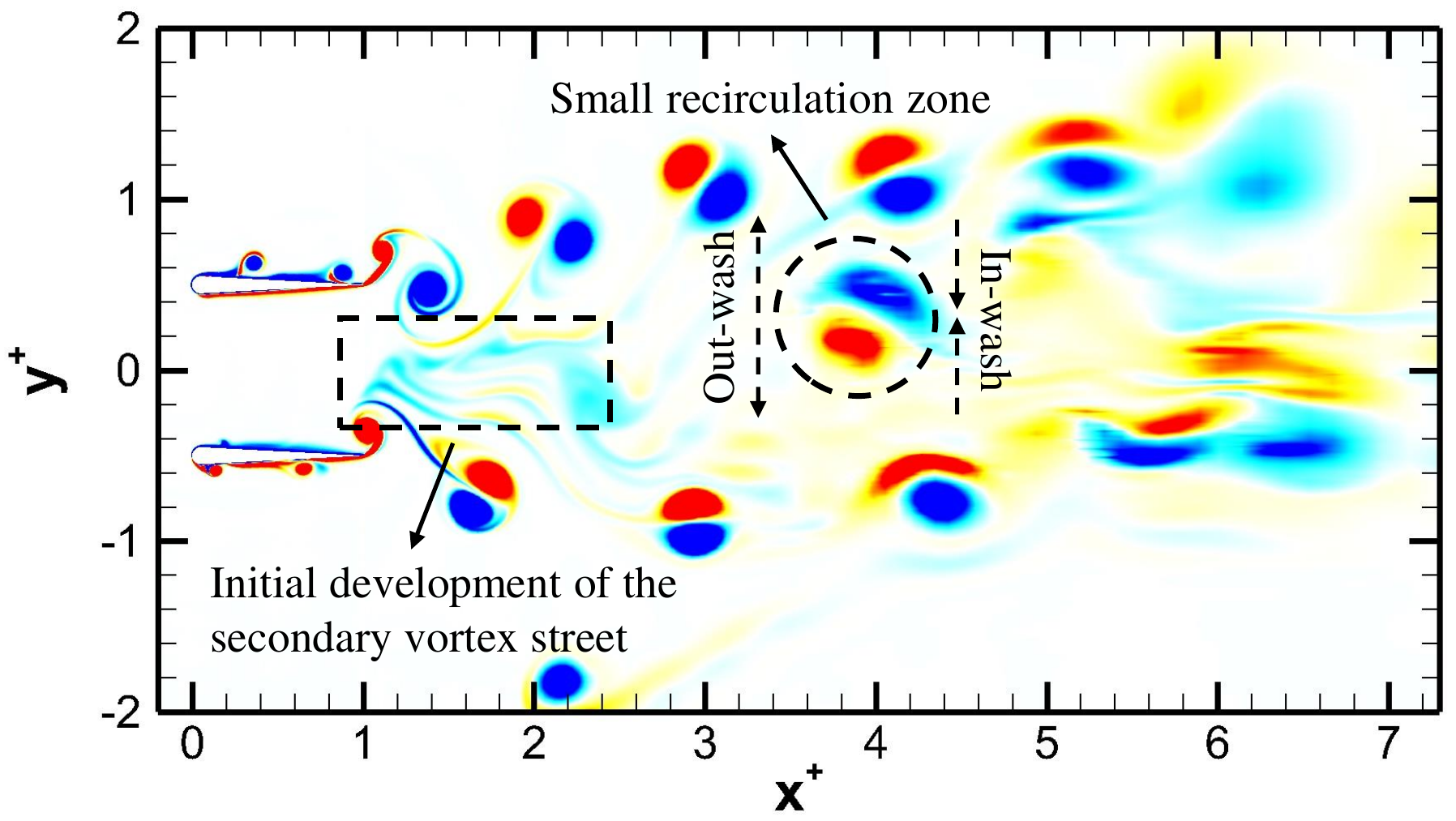}}
\subfigure[$\quad t_6=28T$]{\includegraphics[width=0.47\textwidth]{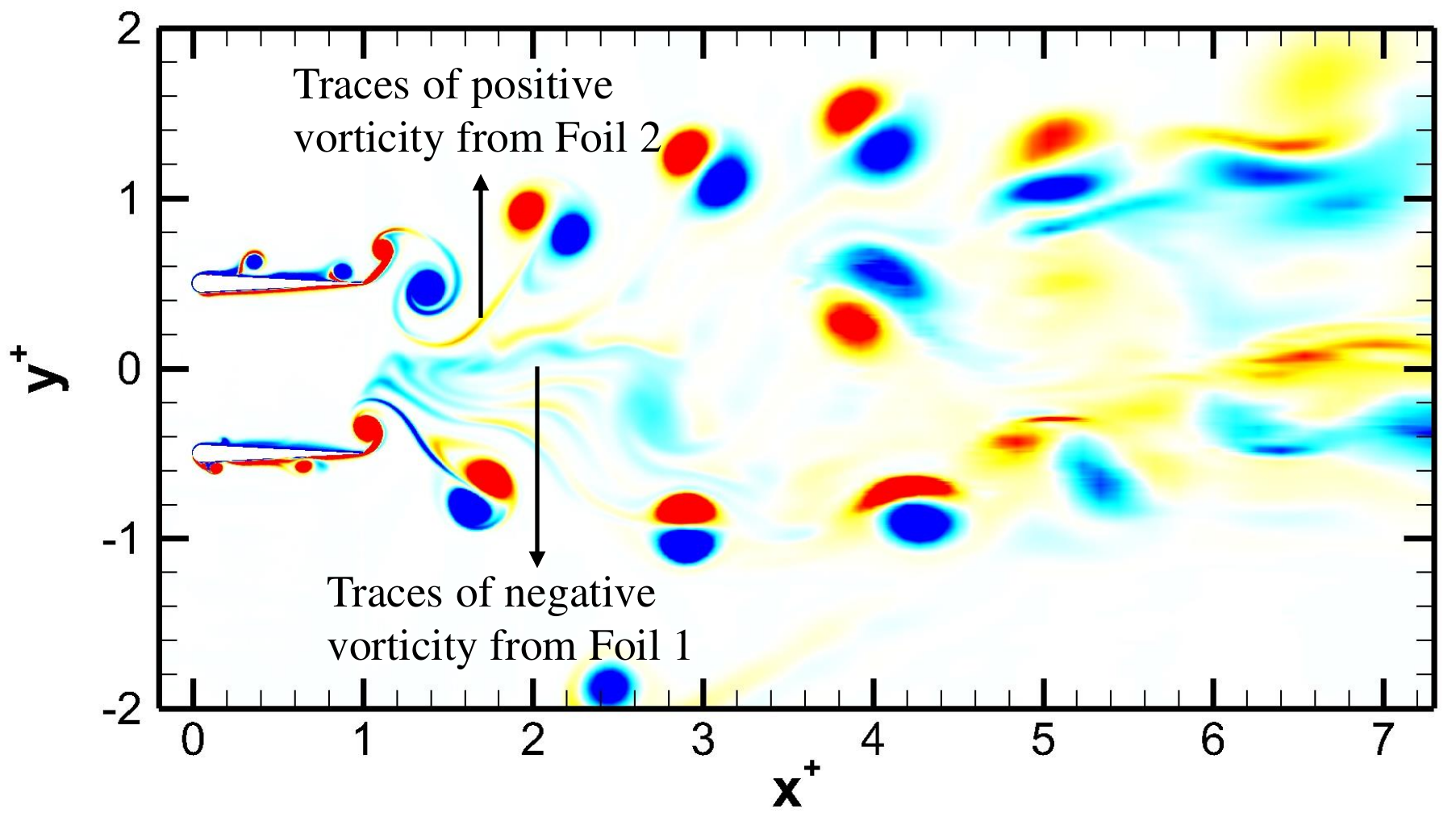}}
\subfigure[$\quad t_7=40T$]{\includegraphics[width=0.47\textwidth]{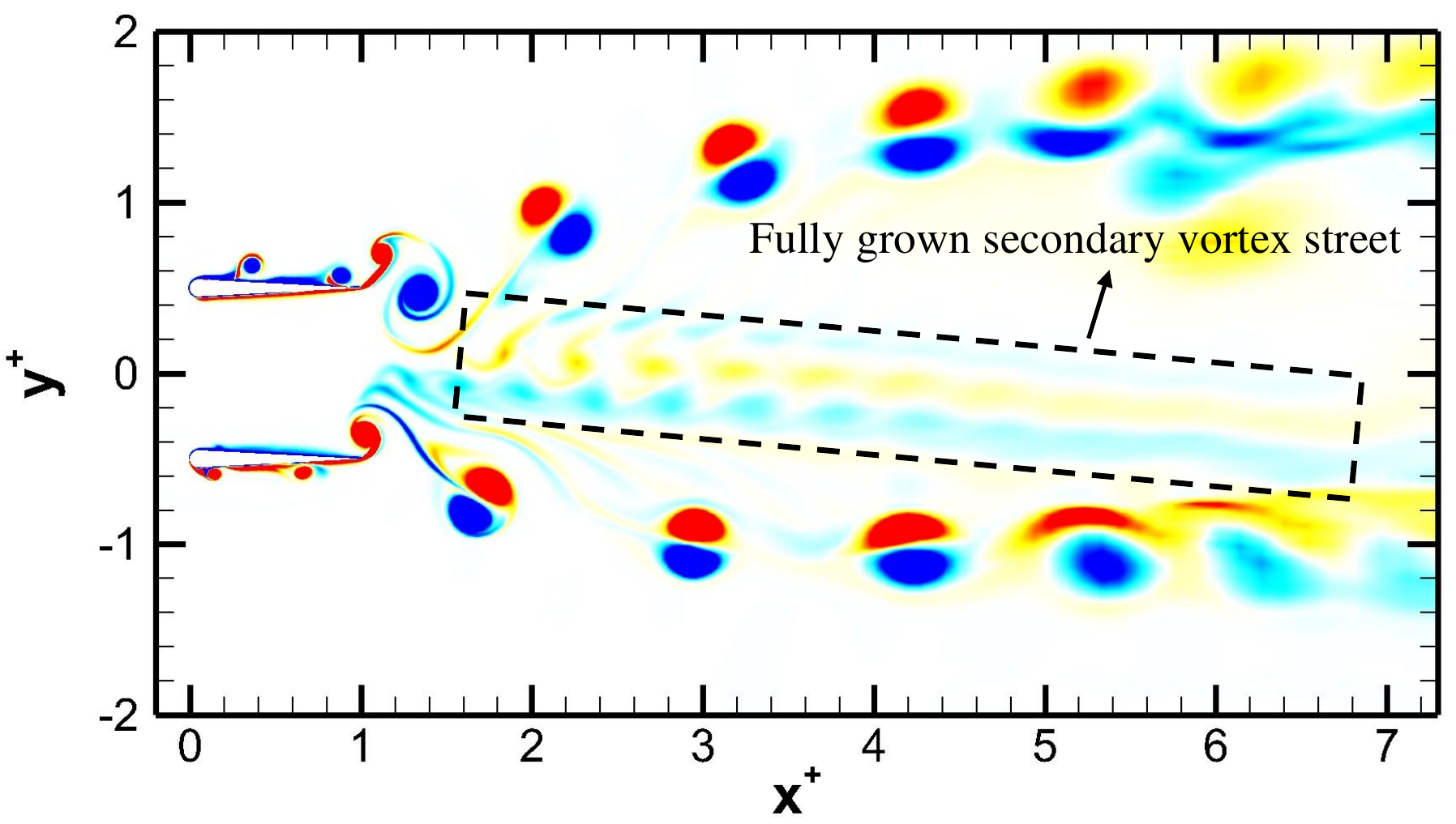}}}	
\caption{Contour of instantaneous spanwise vorticity ($\omega_z^*$) over different pitching cycles for side-by-side foils following Mode 1. Vorticity is normalized by $U_\infty/c$.\\
}
\label{fig_wake-Mode1}
\end{figure}

\subsubsection{Mode 1\\}
Foils began their in-phase oscillations for Mode 1 with an upstroke motion during the $20$\textsuperscript{th} cycle (see supplementary movie 2). At all time instants, two distinct small vortices rotating clockwise (negative vorticity) were identified traveling on the upper surface of Foil 2. On the contrary, two counter-clockwise rotating vortices (positive vorticity) were observed on the lower surface of Foil 1. These vortices were formed at their respective sides of leading edges of the foils, termed as leading edge vortices (\mbox{LEVs}) in Fig.~\ref{fig_wake-Mode1}a. Each foil produced one $\mbox{LEV}$ during a pitching cycle. These vortices separated from the foils only to be merged with structures formed at the trailing edges, which are referred to as trailing edge vortices (\mbox{TEVs}). They were then shed into the wake. It is important to highlight that no distinct coherent structures were formed at the lower side of the leading edge of Foil 2, which proceeded to pass onto the lower surface trailing edges of Foil 2. There were also similar observations on the lack of structures on the upper side of Foil 1. The presence of these foils in the vicinity of each other inhibited the formation of structures on the interior sides of the foils, the side that was closest to the other foil. This hints that  shear layers formed on the interior sides of the two foils may have rolled up due to an induced motion from the foils oscillations, or their interactions, to form $\mbox{TEVs}$ and to pair with their counter rotating vortices shed in the wake.

Figure~\ref{fig_wake-Mode1}a showed the alignment of vortices at the onset of switch in phase difference, that is $t=20.5\tau$. At this instant, two vortex dipoles were observed, represented as $D_{u_1}$ and $D_{l_1}$, moving downstream. These two dipole structures were the last vortex pairs formed by the initial out-of-phase motion, and they did not appear symmetric about the axis represented by $y^+ = 0$. The dipole $D_{l_1}$ traveled upwards due to its higher cross-stream velocity, whereas $D_{u_1}$ exhibited a very small cross-stream displacement. The dipole $D_{u_1}$ had a higher stream-wise convective speed as shown in Fig.~\ref{fig_wake-Mode1}b. During the $22$\textsuperscript{nd} pitching cycle, $D_{l_2}$ began to traverse upstream to interact with the consecutive pair shed from Foil 1. One of the poles of $D_{l_2}$ that had a negative vorticity (blue color in Fig. 4c) pairs with a pole of $D_{l_3}$ with opposite sign vorticity (positive - red color). This formed two new dipoles that were separated and move in opposite directions, one moving upstream towards the foil and another downstream into the wake (see arrows in Figure 4c). The newly formed dipole structure moving upstream remained under the lower primary vortex street and  diffused after a few more oscillation cycles without contributing more to the wake dynamics. After another oscillation cycle, this exchange of pairs and the insensitivity of the downward moving vortex dipole changed the phase of the flow dynamics in the near-wake region. Before the change of the phase angle, the vortices in the lower street had a higher convective speed than those in the upper street, as shown by the longitudinal positions of $D_{u_1}$ and $D_{l_1}$ at the end of the $21$\textsuperscript{st} pitching cycle in Figs.~\ref{fig_wake-Mode1}a and \ref{fig_wake-Mode1}b.
 
After the $22$\textsuperscript{nd} cycle, vortex dipoles shed from Foil 2 in the upper vortex street led those shed from Foil 1. Newly formed $D_{l_2}$ and $D_{u_2}$ approached each other until their initial interactions in the $24$\textsuperscript{th} cycle formed a small recriculatory zone, which was identified by the dashed rectangle in Fig.~\ref{fig_wake-Mode1}e. Two vortices in the middle of this region retained their position, hence forming a small region of consistent recirculatory fluid, and the other two were convected downstream. The newly formed recirculary flow region between the two vortex streets remained intact for $4$ cycles, until the end of the $28$\textsuperscript{th} cycle in Fig.~\ref{fig_wake-Mode1}g. Although this local recirculatry region moved downstream, its convective speed seems very low. Moreover, the circulatory motion of the fluid inside this region induced secondary upwash and downwash flow after and prior to this zone, which were identified in Fig.~\ref{fig_wake-Mode1}f at $x^+=4$. This intermediate dynamics provided some additional space for a secondary vortex street to be formed by the traces of positive vorticity from the wake of Foil 2 and negative vorticity from that of Foil 1, highlighted in the rectangular region in Fig.~\ref{fig_wake-Mode1}h. On the contrary, \cite{raj2020jet} found the jet deflecting on one side due to the interaction of vortices shed by their Foils 1 and 2 in a similar geometric configuration but oscillating in a quiescent fluid. For the remaining oscillation cycles, this secondary vortex street grew, the induced flow from which kept the two primary streets from merging with each other. It is important to note that the vortex pair in the recirculation zone, shown in Fig.~\ref{fig_wake-Mode1}f, stayed in the middle of the two primary vortex streets only to diffuse at a later stage by the induced momentum from the newly forming secondary vortex street. Hence, the wake did not merge to form a single vortex street similar to that for a pure in-phase pitching case \citep{gungor-PRE-2020}. The same wake mechanism was also observed for Mode 2.

\subsubsection{Mode 2\\}
\begin{figure}
\centering{\subfigure{\includegraphics[width=0.35\textwidth]{vorticity_legend.pdf}}}\\
\subfigure[$\quad t_0=21\tau$]{\includegraphics[width=0.47\textwidth]{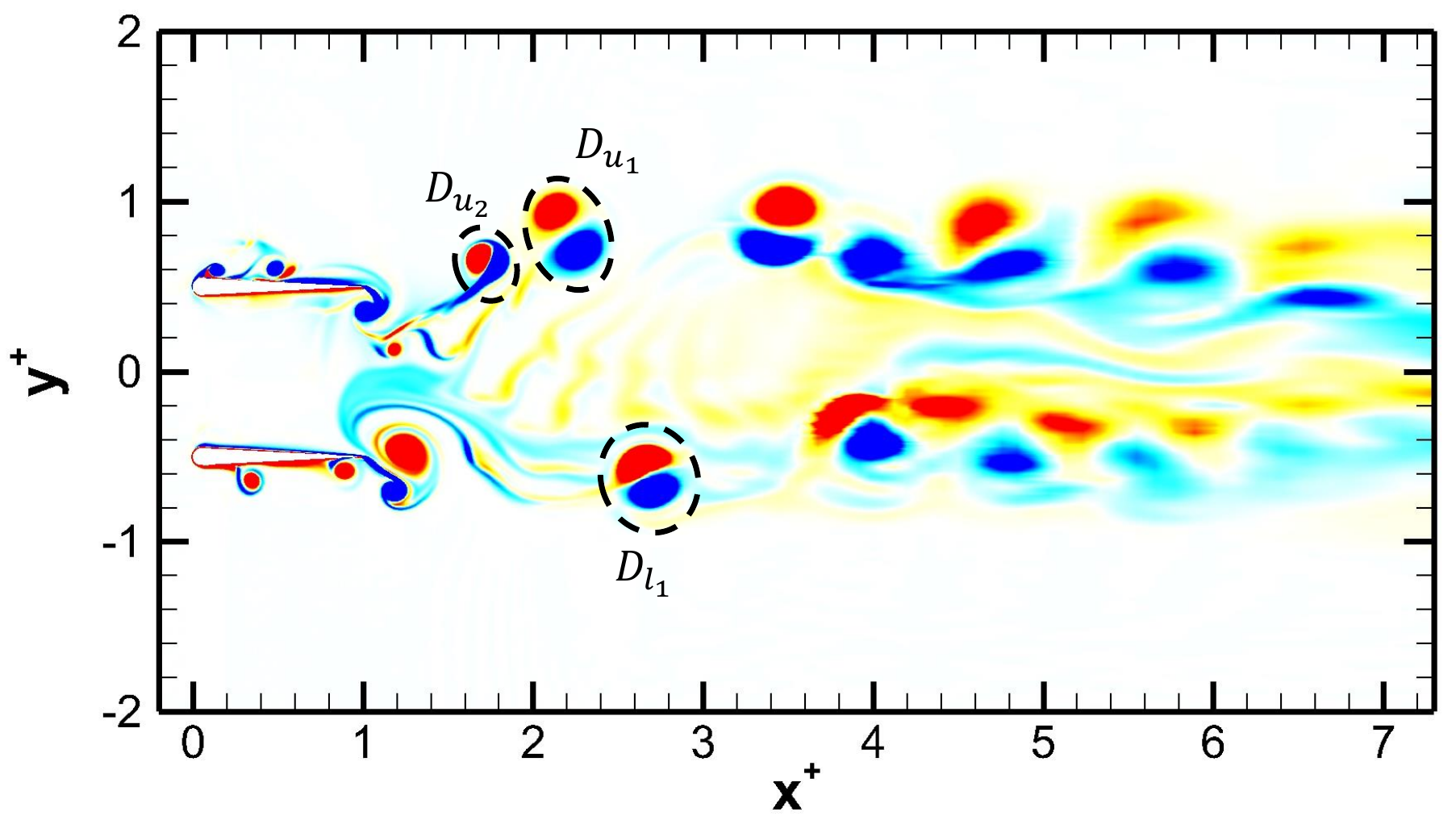}}
\subfigure[$\quad t_1=22\tau$]{\includegraphics[width=0.47\textwidth]{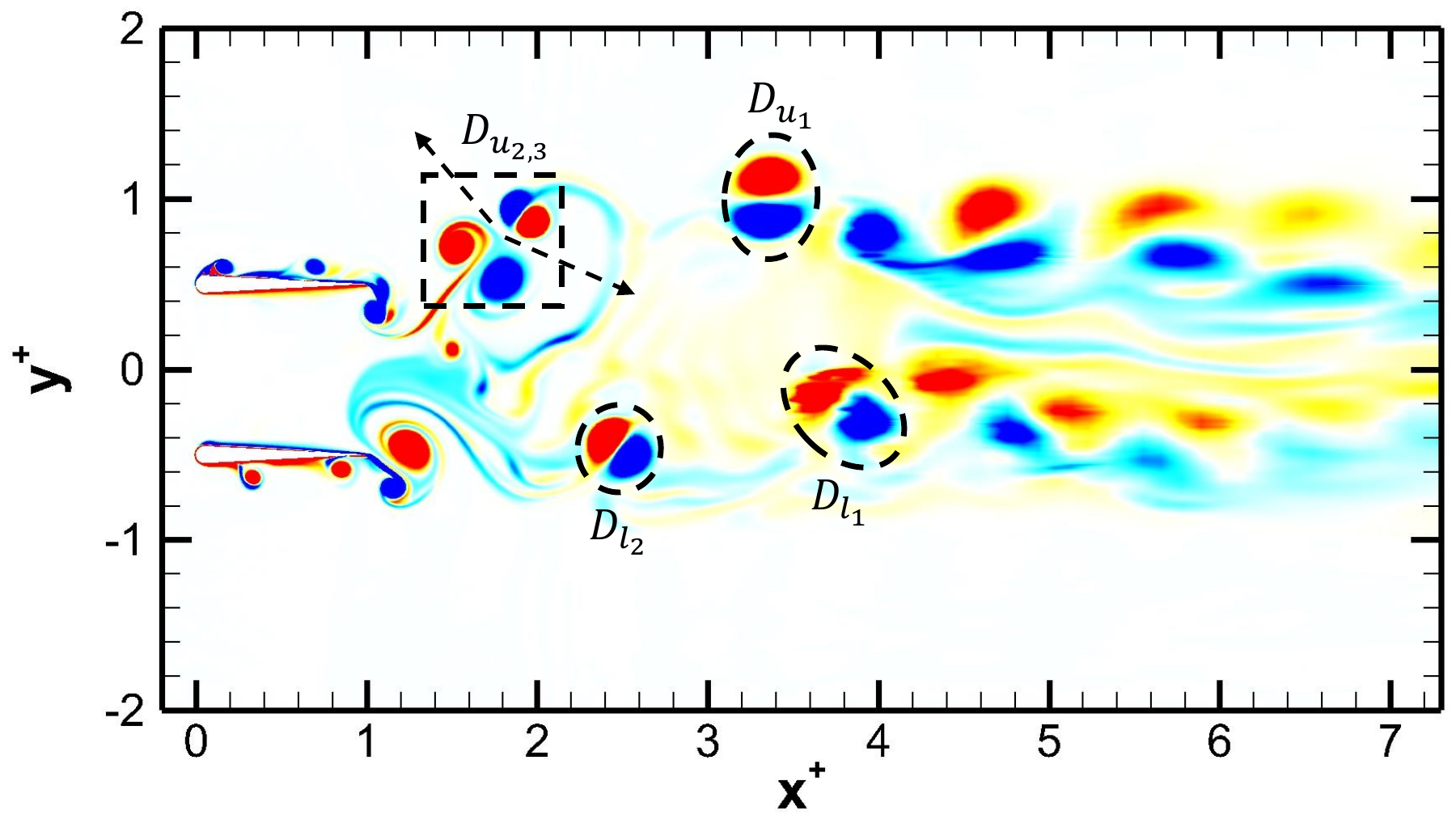}}
\subfigure[$\quad t_2=23\tau$]{\includegraphics[width=0.47\textwidth]{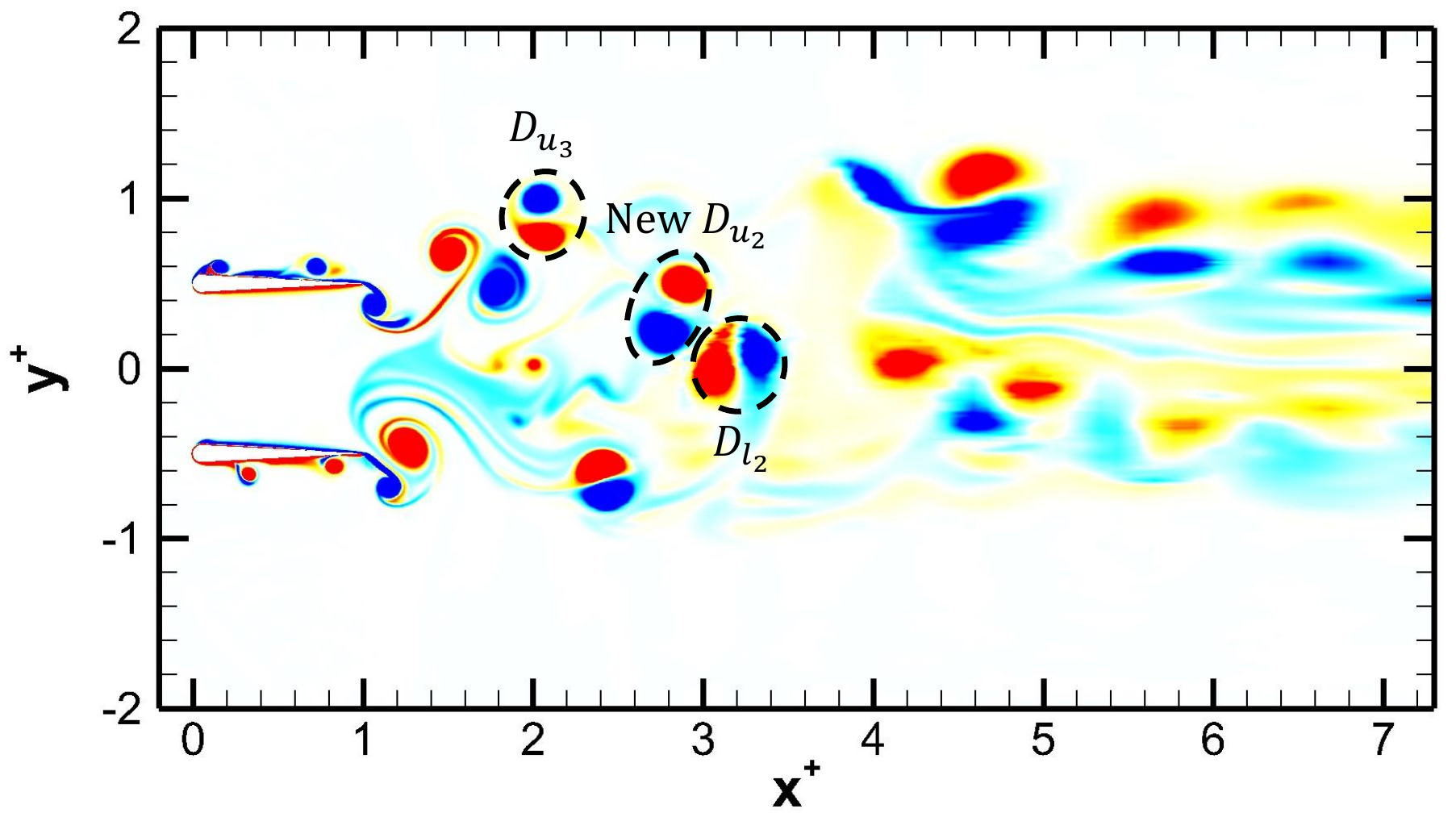}}	
\subfigure[$\quad t_3=24\tau$]{\includegraphics[width=0.47\textwidth]{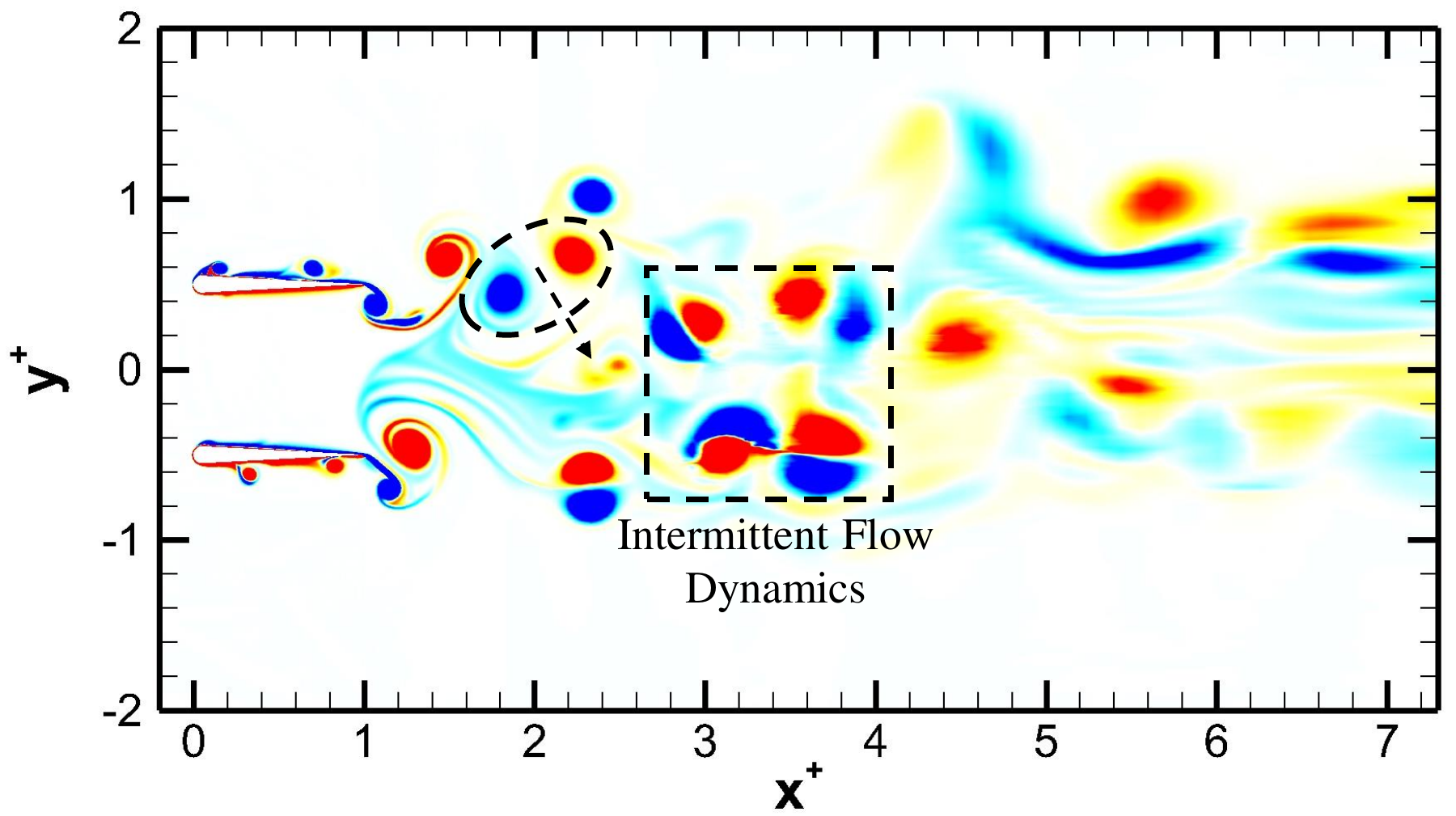}}
\subfigure[$\quad t_4=25\tau$]{\includegraphics[width=0.47\textwidth]{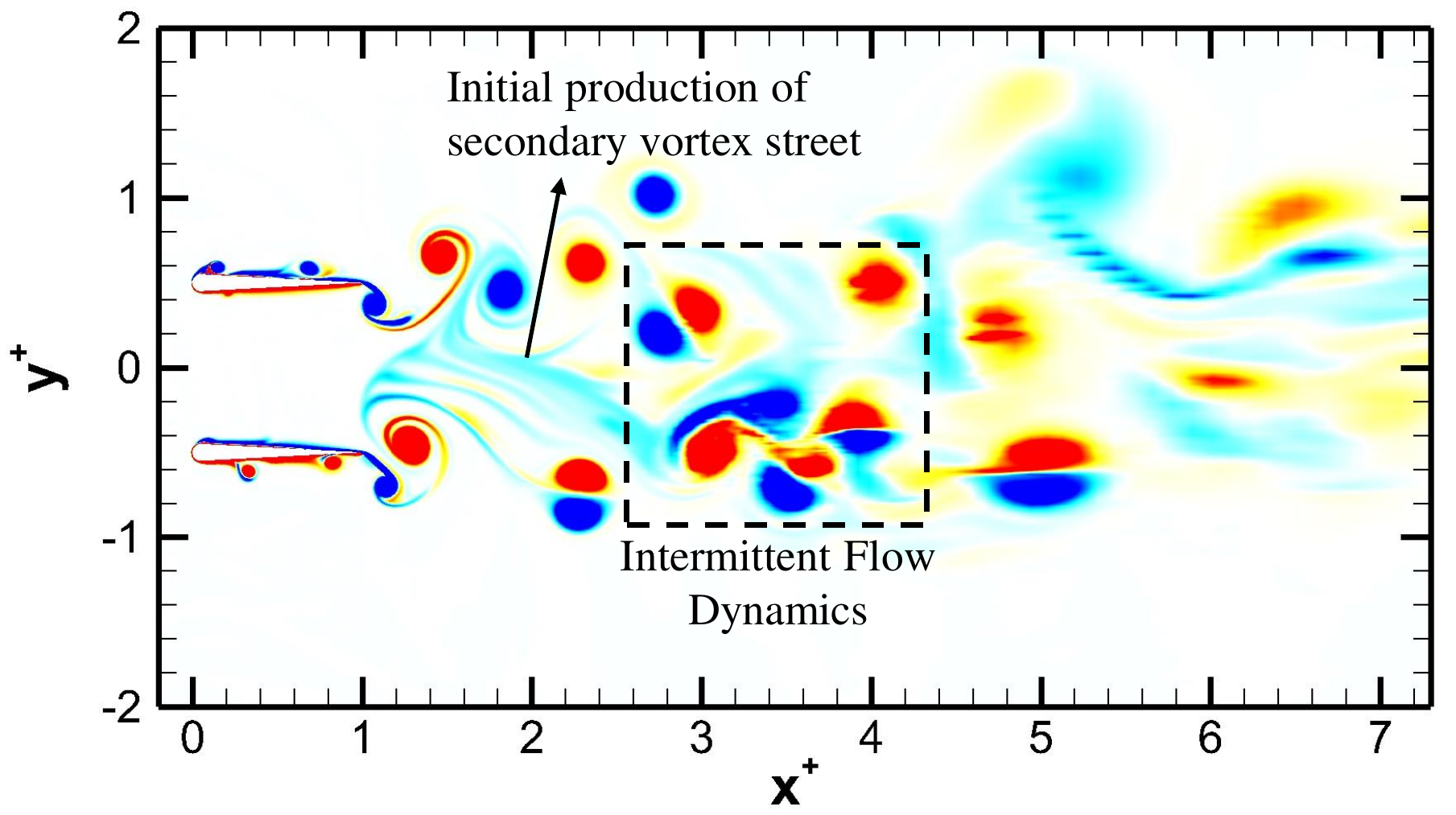}}	
\subfigure[$\quad t_5=27\tau$]{\includegraphics[width=0.47\textwidth]{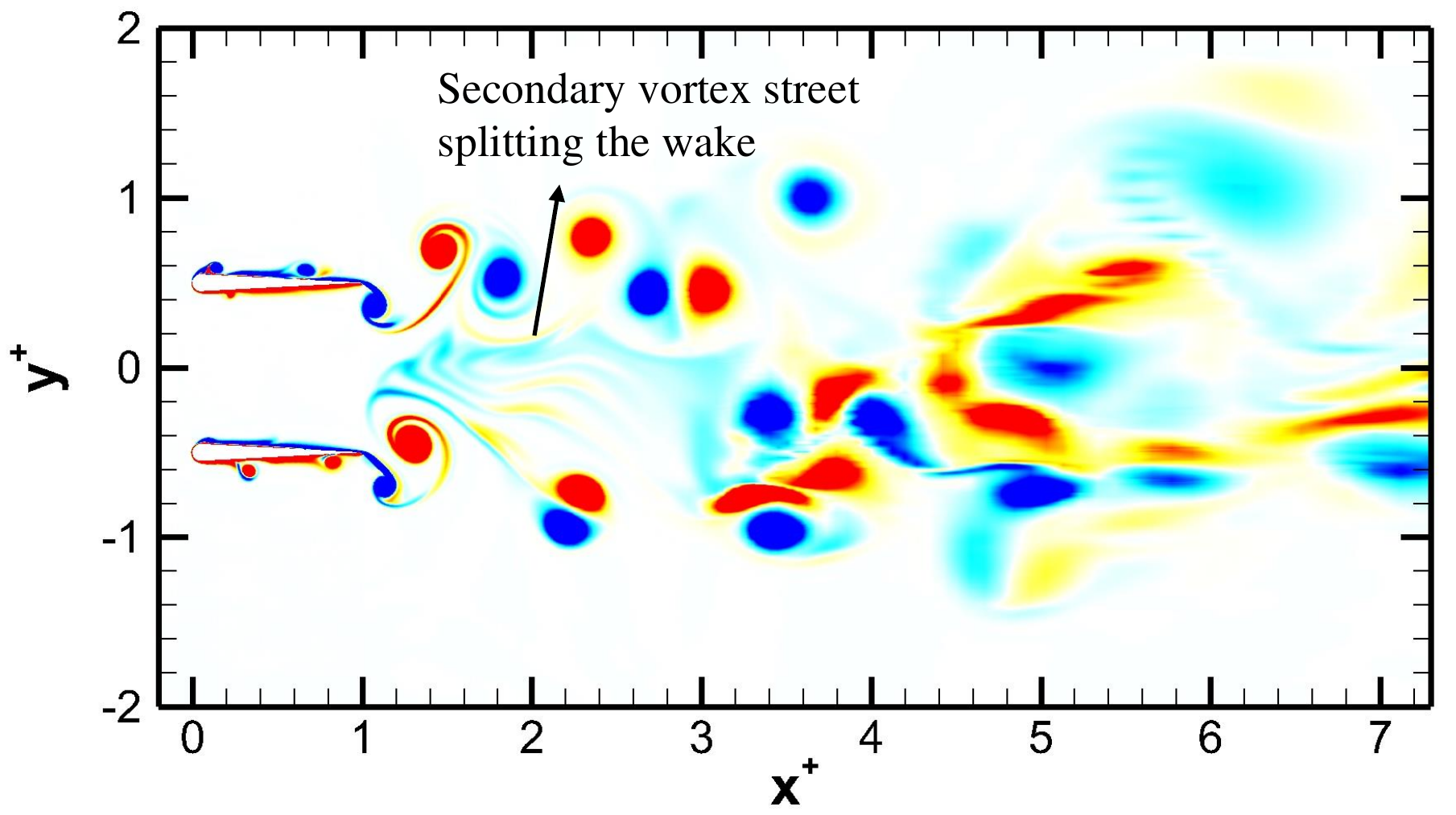}}
\subfigure[$\quad t_6=32\tau$]{\includegraphics[width=0.47\textwidth]{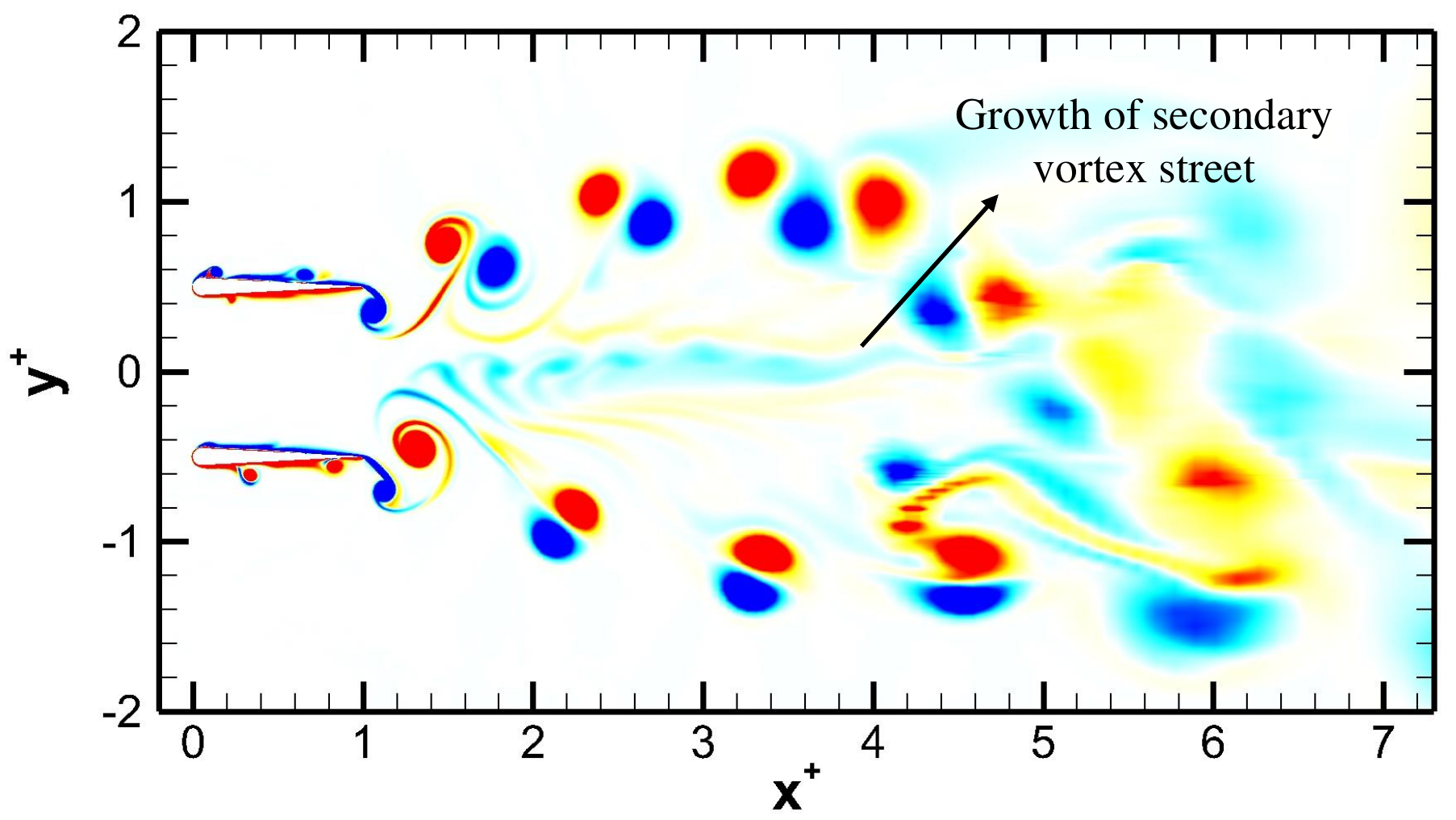}}	
\hspace{7mm} 
\subfigure[$\quad t_7=40\tau$]{\includegraphics[width=0.47\textwidth]{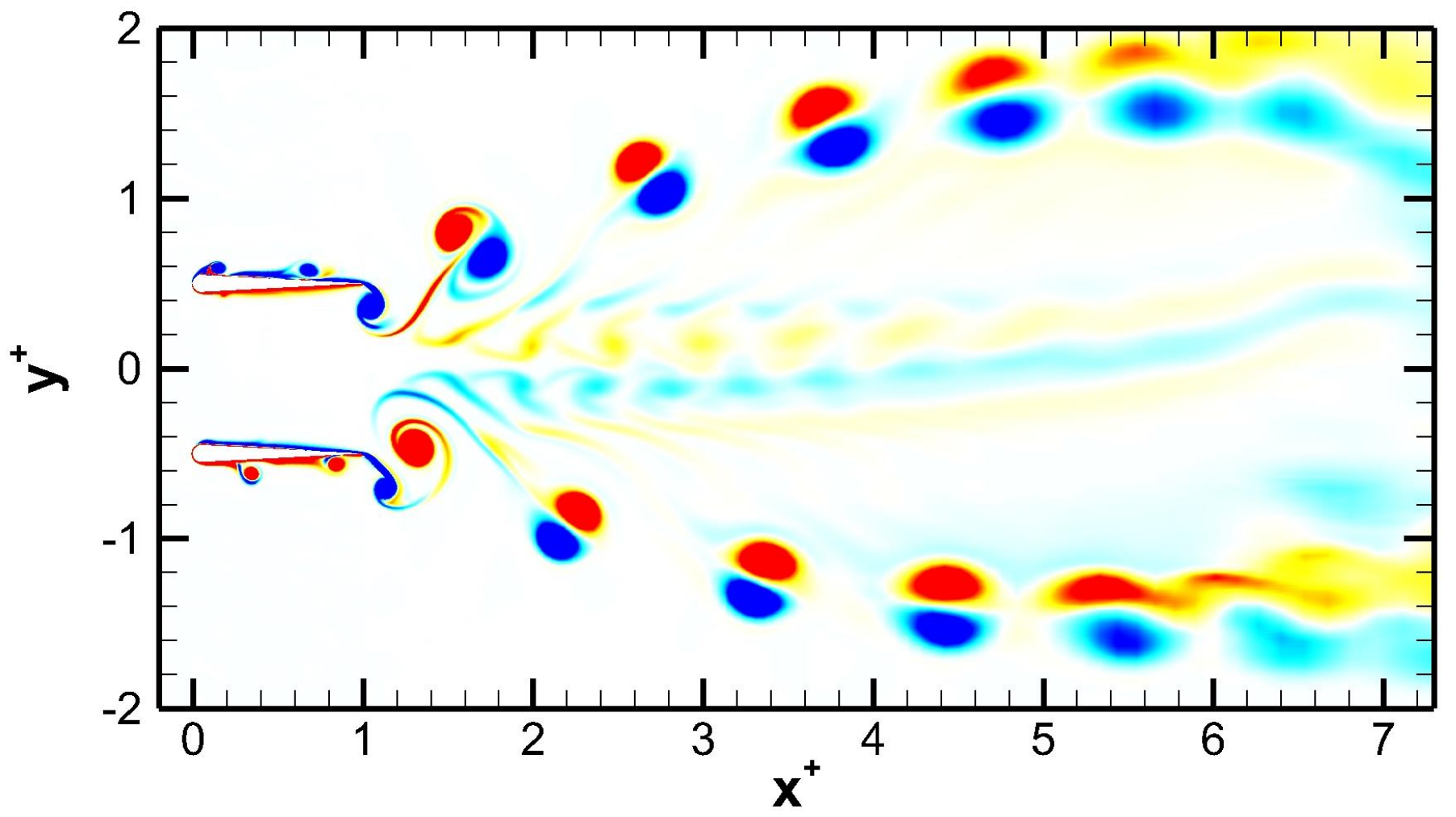}}			
\caption{Contour of instantaneous spanwise vorticity ($\omega_z^*$) over different pitching cycles for side-by-side foils following Mode 2. The contour values are the same as in Fig. \ref{fig_wake-Mode1}.}
\label{fig_wake-Mode2}
\end{figure}

The configuration and oscillations of the foils in Mode 2 were similar to those of Mode 1, but the initial direction of pitching was reversed following the change of the phase angle during the $20$\textsuperscript{th} cycle. Thus, the wake shown in Fig.~\ref{fig_wake-Mode1}a remained the same for Mode 2. In the following half cycle, i.e., at the end of the $21$\textsuperscript{st} pitching cycle, a vortex dipole ($D_{u_2}$ in Fig.~\ref{fig_wake-Mode2}a) was observed in the wake of Foil 2, whereas its counterpart vortex dipole was shed by Foil 1 with a lag. Moreover, the leg of one pole of $D_{u_2}$ remained connected to the foil's trailing edge. This lag dictated the wake dynamics for the next few oscillation cycles. At the end of the $22$\textsuperscript{nd} cycle in Fig.~\ref{fig_wake-Mode2}b, $D_{u_2}$ retained its position, which followed by its interactions with the newly forming dipole, $D_{u_3}$. The poles with opposite sign vorticity in each structure paired (negative pole of $D_{u_2}$ with the positive pole of $D_{u_3}$) to form two new dipoles that move in opposite directions. This dynamics was attributed to the impulsive motion associated with the collision of two wake structures. Such a symmetry-breaking phenomenon was also observed by \cite{bao2017dynamic} due to strong flow mixing in the wake of two parallel foils performing out-of-phase pitching oscillations at a chord-based $\mbox{Re}=1275$. This interactive process formed two very different structures, one that interjected with the wake development by moving towards the center of the wake (that is ``New $D_{u_3}$" moving towards $y^+=0$ in Fig.~\ref{fig_wake-Mode2}c), and one that moved away from the wake (see $D_{u_3}$ in Fig.~\ref{fig_wake-Mode2}c). The interaction of New $D_{u_2}$ with $D_{l_2}$ at the end of the $23$\textsuperscript{rd} cycle (see Fig.~\ref{fig_wake-Mode2}c) formed the region of intermittent vortex interaction between $x^{+}=2.8$ and $x^{+}=4$. This process continued with the consecutive vortex pairs formed by Foil 2 entering this region when the $24$\textsuperscript{th} cycle was completed (see Fig.~\ref{fig_wake-Mode2}d). This process enabled the formation of a secondary vortex street portrayed in Fig.~\ref{fig_wake-Mode2}e. At this stage, this formation was dominated by small structures with negative vorticity from both foils. From the $27$\textsuperscript{th} pitching cycle shown in Fig.~\ref{fig_wake-Mode2}f, the secondary vortex street grew downstream and became more prominent. Quite interestingly, this expanding vortex street split the overall wake by pushing the intermittently interacting coherent structures further downstream. Although there were no upwash and downwash induced flow in the wake due to the intermittent vortex interaction zone, the presence of this circulatory flow region deflected the primary vortex streets. Thus, vortex structures moved downstream in a deflected wake for their respective foil without any apparent interactions, hinting that their dynamics appear not to depend on one-another. As the secondary vortex grew past the $32$\textsuperscript{nd} cycle in Fig.~\ref{fig_wake-Mode2}g, the wake deflections persisted through out the remaining cycles of oscillation, approaching a quasi-steady behavior. This aligned with the variations in thrust, power and side-force.

\subsubsection{Mode 3\\}
\begin{figure}
\centering{\subfigure{\includegraphics[width=0.35\textwidth]{vorticity_legend.pdf}}}\\
\subfigure[$\quad t_0=22.5\tau$]{\includegraphics[width=0.47\textwidth]{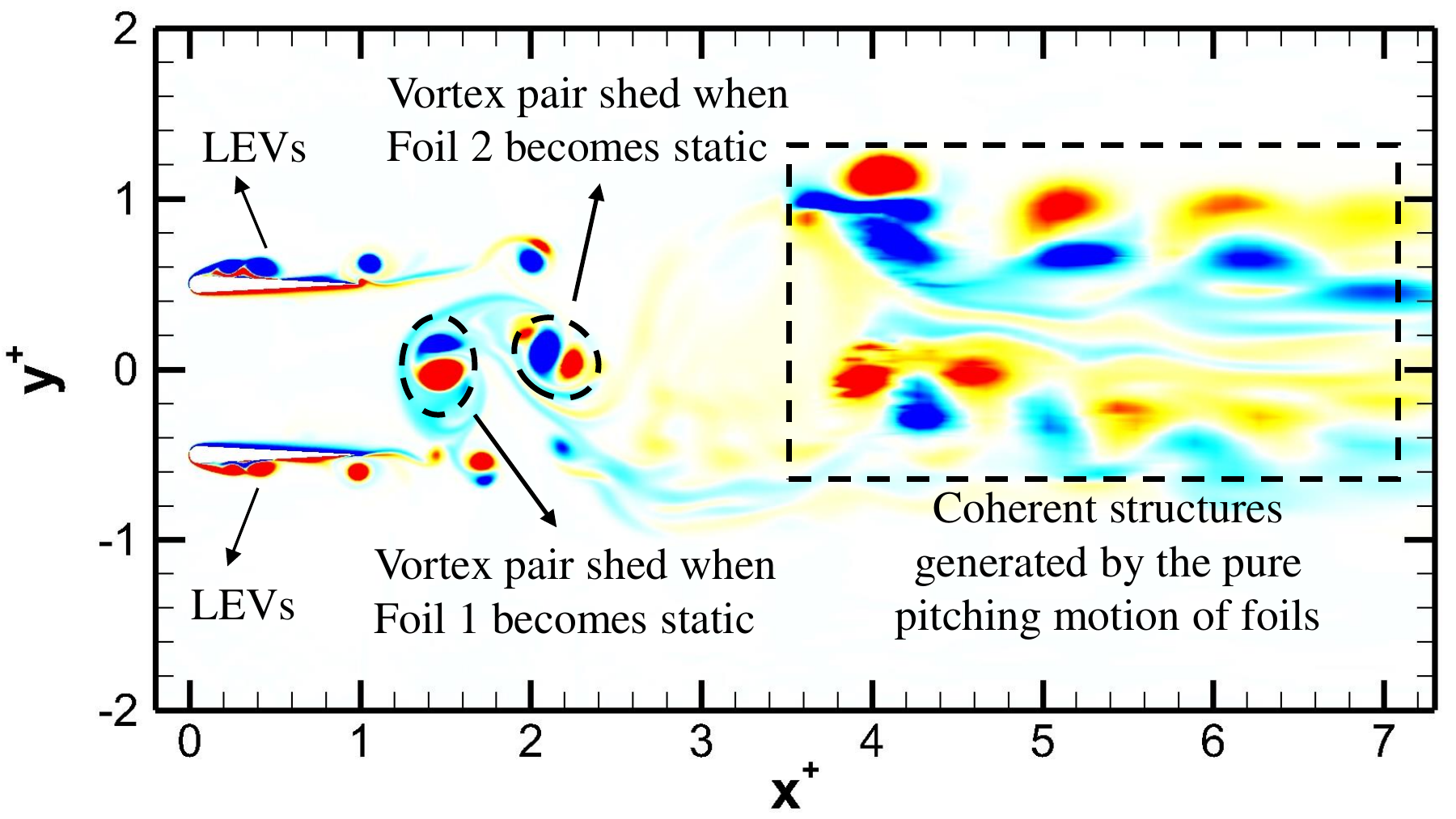}}
\subfigure[$\quad t_1=25\tau$]{\includegraphics[width=0.45\textwidth]{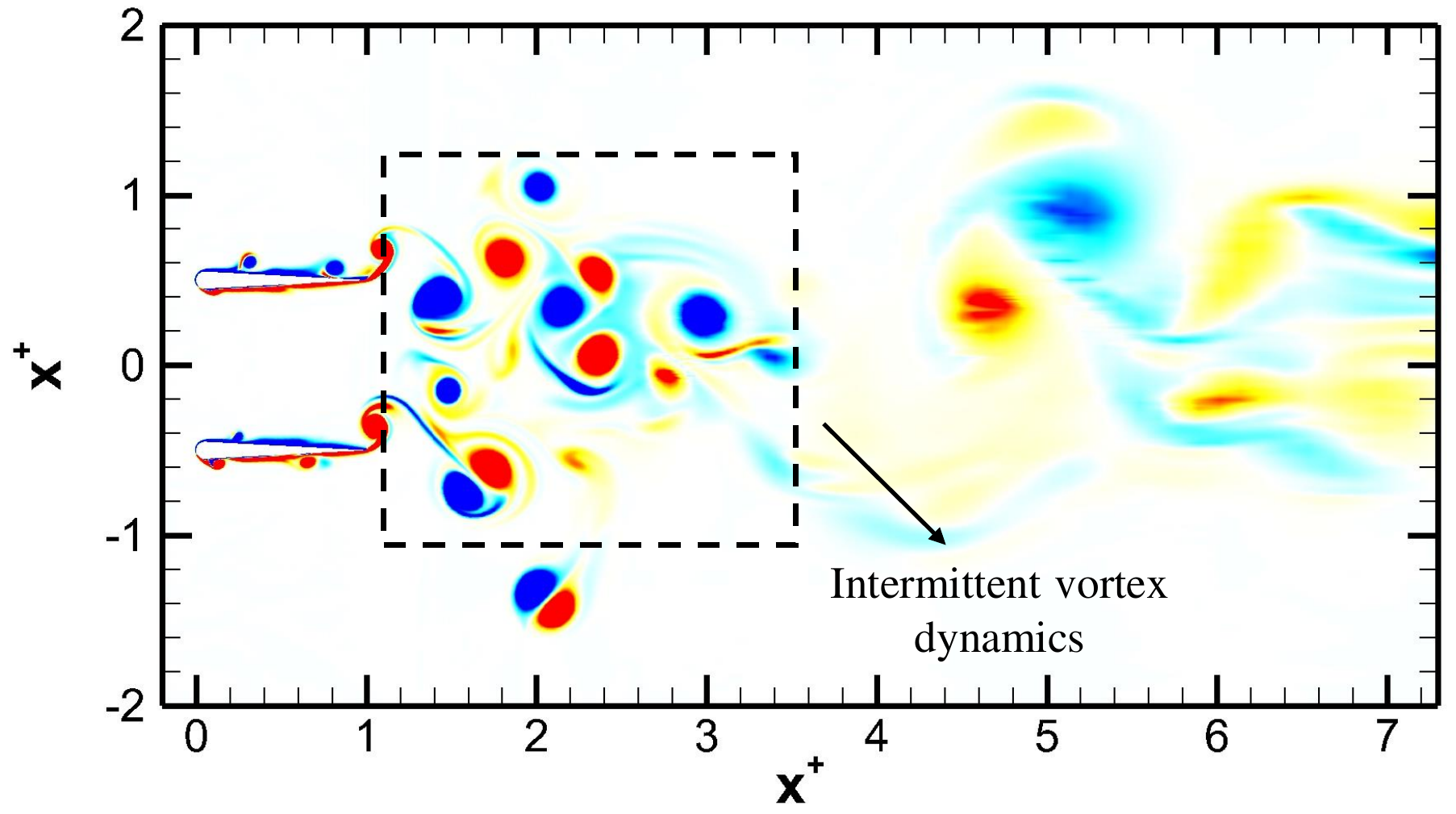}}
\subfigure[$\quad t_2=28\tau$]{\includegraphics[width=0.47\textwidth]{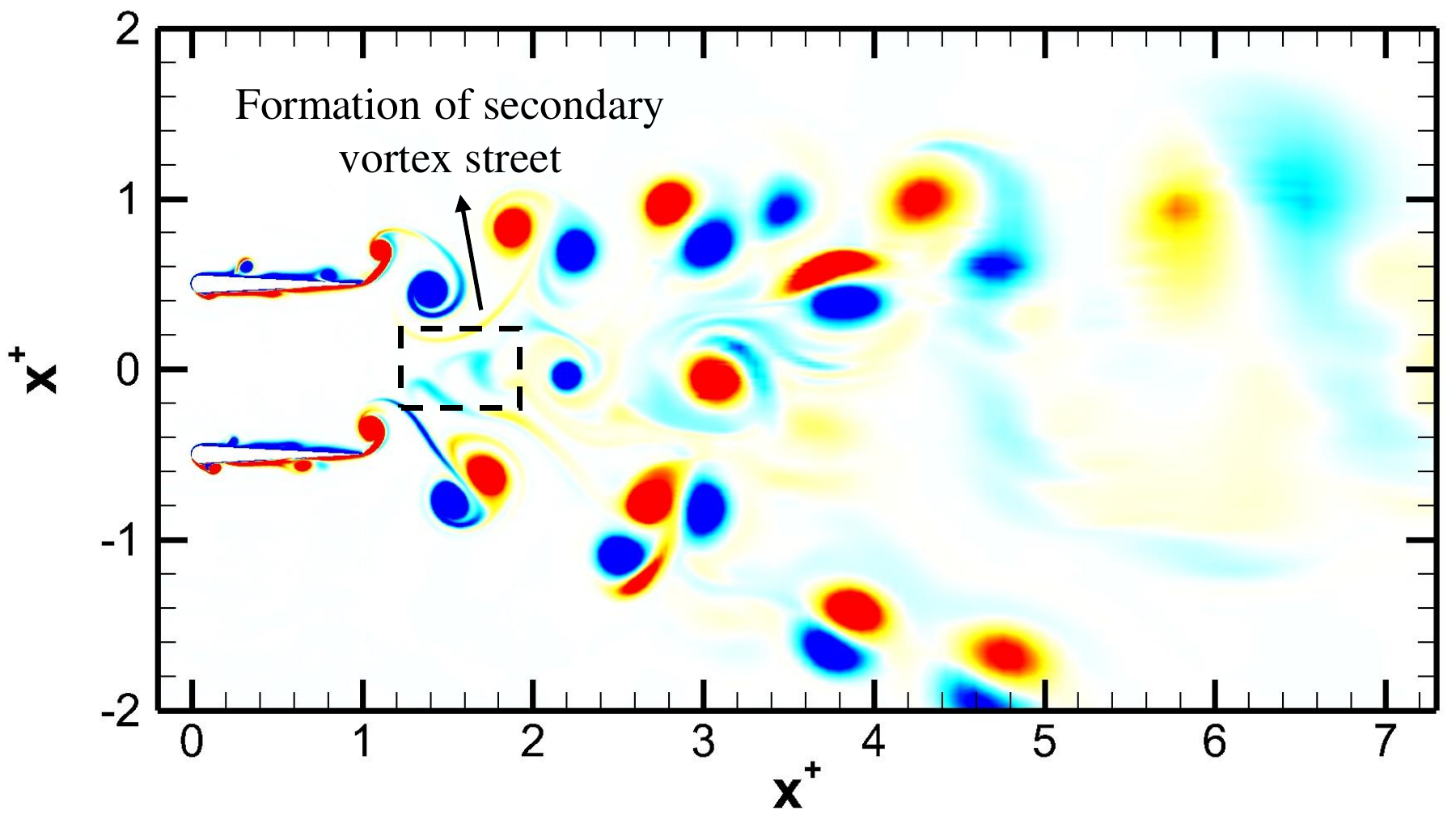}}	
\subfigure[$\quad t_3=32\tau$]{\includegraphics[width=0.47\textwidth]{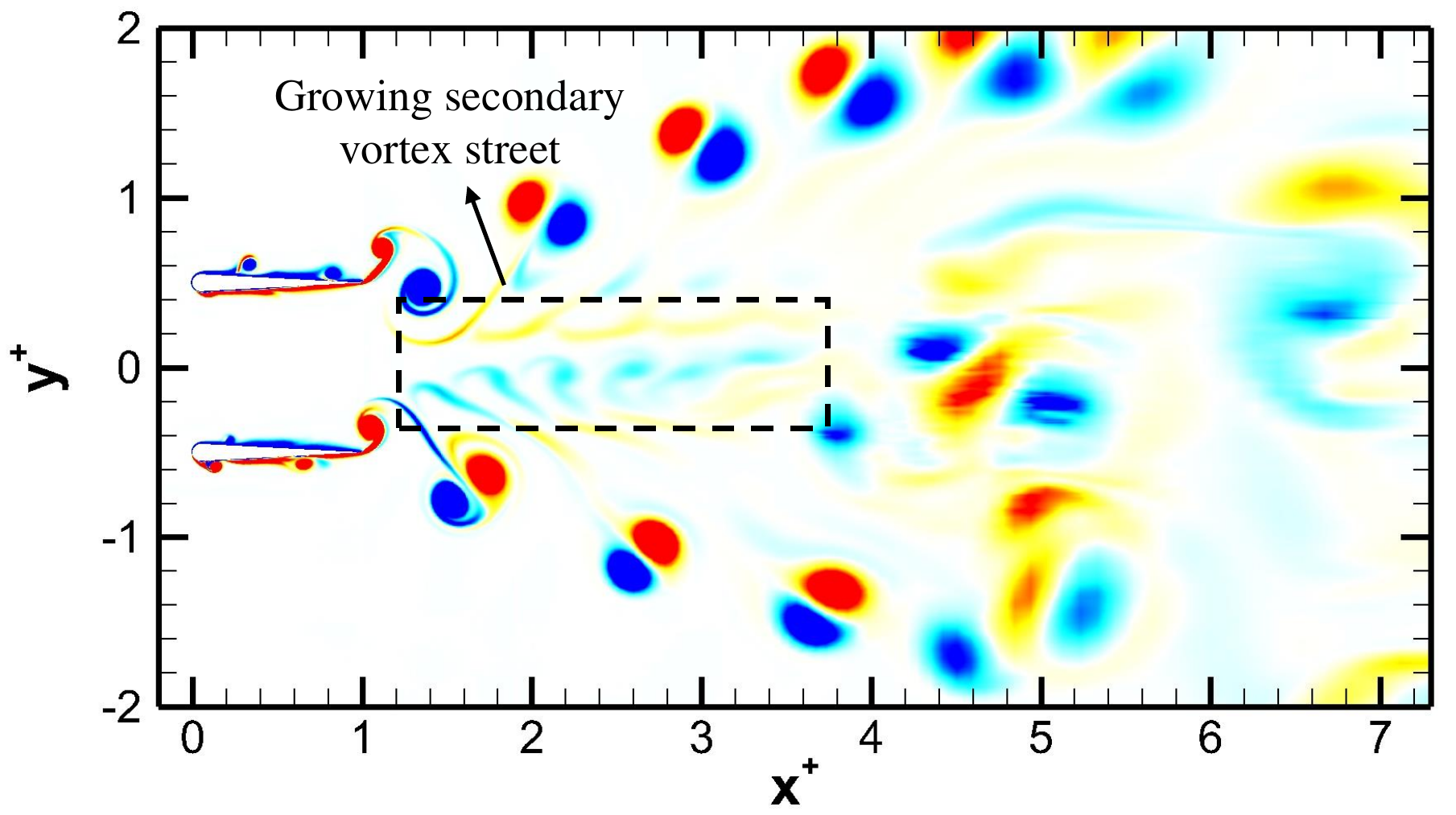}}
\subfigure[$\quad t_4=36\tau$]{\includegraphics[width=0.47\textwidth]{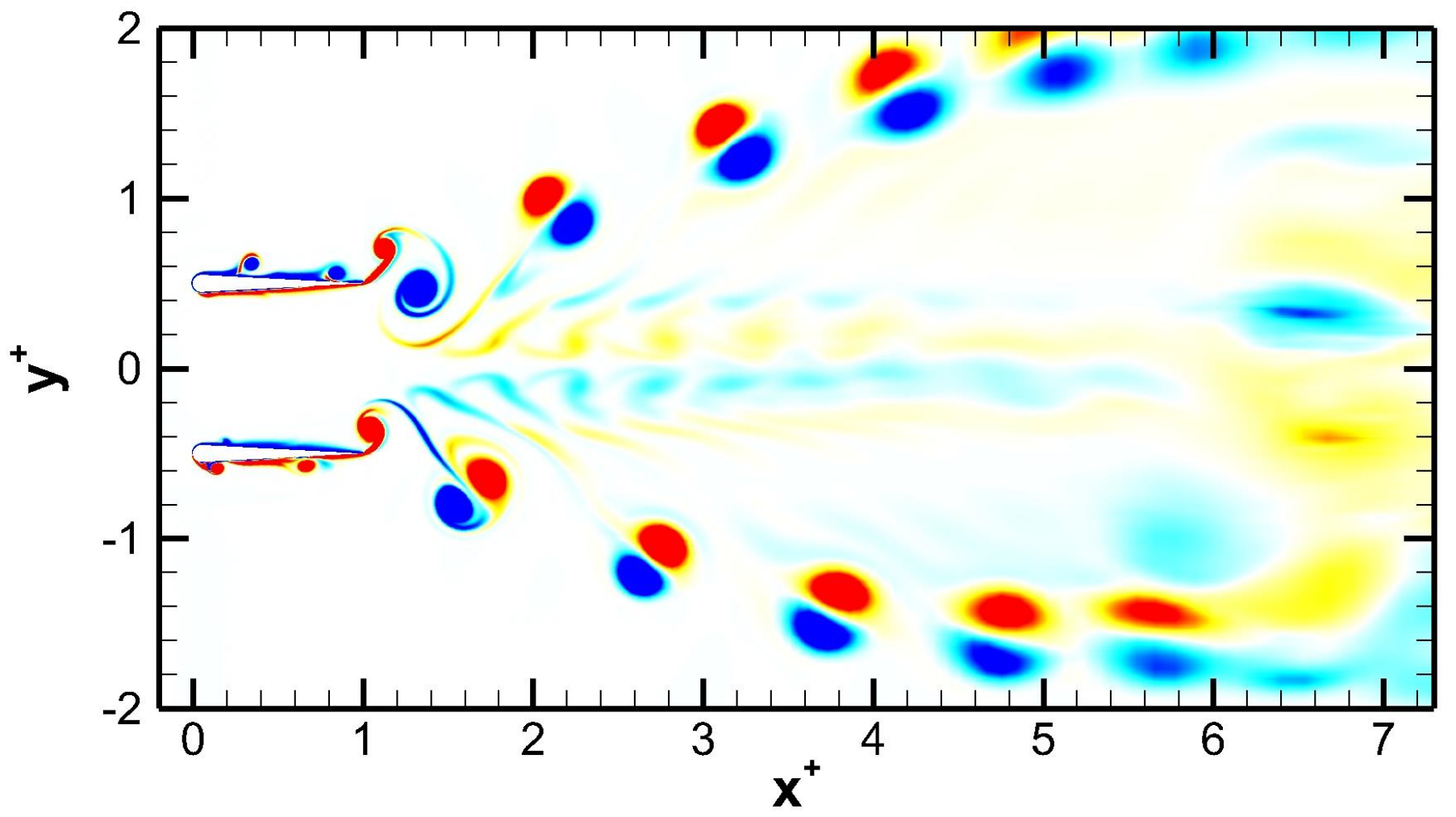}}	
\hspace{7mm} 
\subfigure[$\quad t_5=40\tau$]{\includegraphics[width=0.47\textwidth]{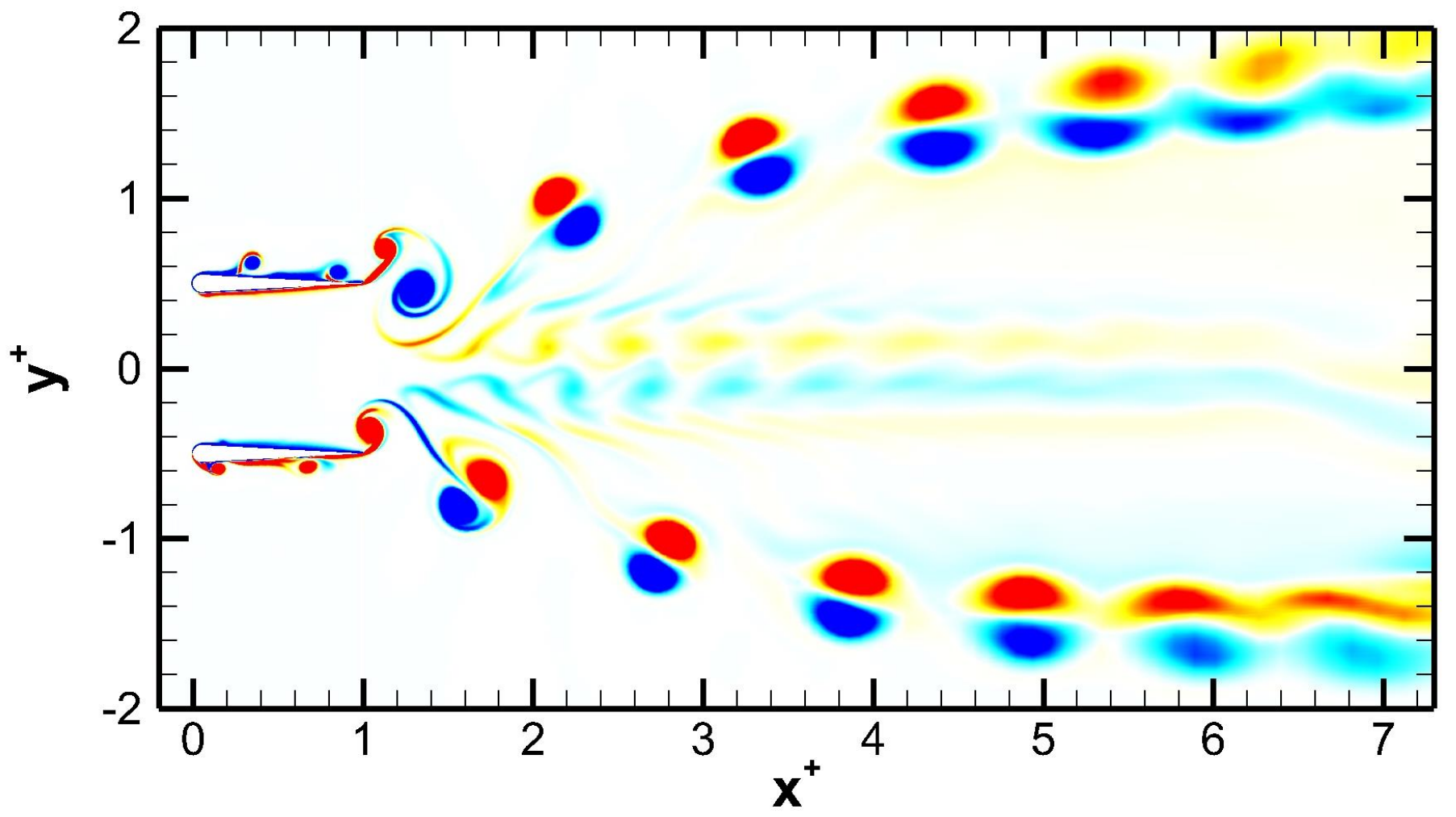}}			
\caption{Contour of instantaneous spanwise vorticity ($\omega_z^*$) over different pitching cycles for side-by-side foils following Mode 3. The contour values are the same as in Fig. \ref{fig_wake-Mode1}.}
\label{fig_wake-Mode3}
\end{figure}

The oscillatory motion of foils in Mode 3 and Mode 4 were suspended for 2 cycles after the 20th cycle, and then they began their in-phase oscillations. Thus, Fig.~\ref{fig_wake-Mode1}a still represents the initial state, at which the foils stopped their out-of-phase pitching. Upon the start of oscillations, following two cycles of foils remaining in the horizontal configuration, two $\mbox{LEVs}$ were formed on the lower and upper surfaces of Foil 1 and Foil 2, respectively (see Fig.~\ref{fig_wake-Mode3}a). No counter rotating vortices were generated by either foils, while there were no vortex-pairing observed near the trailing parts of the foils. Figure~\ref{fig_wake-Mode3}a also shows the traversing of coherent structures produced by the out-of-phase pitching motion of foils beyond $x^+ = 3.4$. 

The static state of foils for some time allowed the wake to redevelop itself near the trailing edges without having interaction with previously shed structures. Figure~\ref{fig_wake-Mode3}b shows contours of the out-of-plain vorticity component at the end of the $25$\textsuperscript{th} cycle. It is evident that there was no interactions inside the region enclosed by $3.5 < x^+ < 4.3$. Simultaneously, the formation and shedding of more vortex pairs for the next three oscillation cycles formed a region with intermittent flow dynamics (see supplementary movie 4) that is identified by a dashed rectangular enclosure in Fig.~\ref{fig_wake-Mode3}b. This region is referred to as an intermittent regime because these vortices underwent intense interactions within this region, and no distinct pairing mechanisms were identified. During three more oscillation periods, two interesting phenomena were  observed: (1) a secondary vortex street, identified by a small box in Fig.~\ref{fig_wake-Mode3}c, was formed by the trailing parts of larger structures with negative vorticity (blue color in Fig.~\ref{fig_wake-Mode3}) that were shed from Foil 1 in the middle of the two primary vortex streets, and (2) a circulatory fluid zone was formed by the vortices shed during the $23$\textsuperscript{rd} to $28$\textsuperscript{th} cycles, which convected downstream. The circulatory zone appeared to exhibit a high convective speed, such that it moved downstream quickly compared to previous observations of similar processes in Mode 1 and Mode 2.

The growth of the secondary vortex street (shown in Fig.~\ref{fig_wake-Mode3}d) appeared to be the key factor in the primary vortex streets attaining their quasi-steady behavior by bifurcating the region of the main flow activity. Evidently, the angles of the primary vortex street did not change after the $30$\textsuperscript{th} cycle, that is after the secondary vortex street grew to dominate the centre of the wake. To this effect, the dynamics of these vortex streets can be said to approach a quasi-steady nature after the $36$\textsuperscript{th} cycle.

\subsubsection{Mode 4\\}
\begin{figure}
\centering{\subfigure{\includegraphics[width=0.35\textwidth]{vorticity_legend.pdf}}}\\
\subfigure[$\quad t_1=23\tau$]{\includegraphics[width=0.47\textwidth]{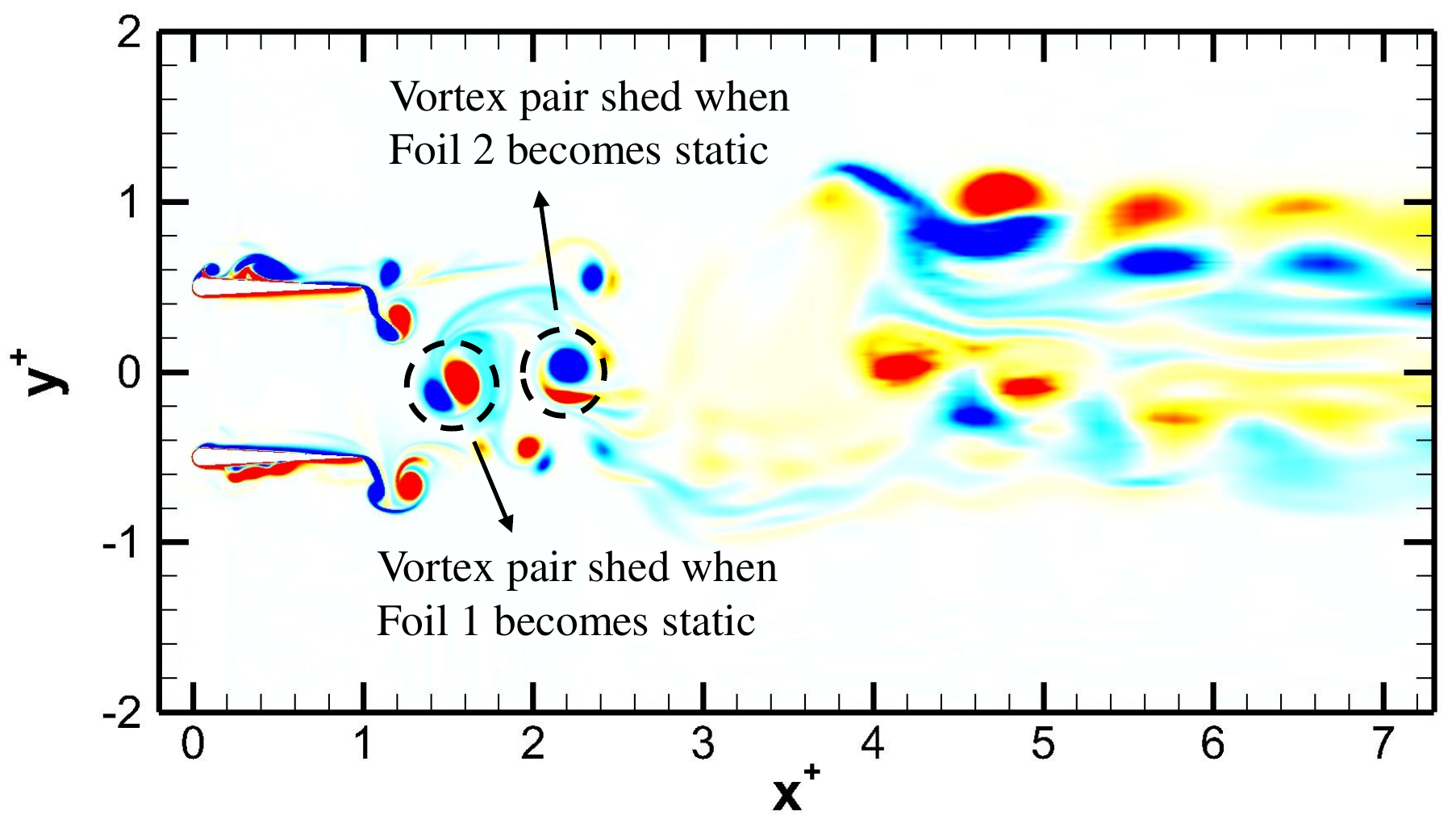}}
\subfigure[$\quad t_2=24\tau$]{\includegraphics[width=0.47\textwidth]{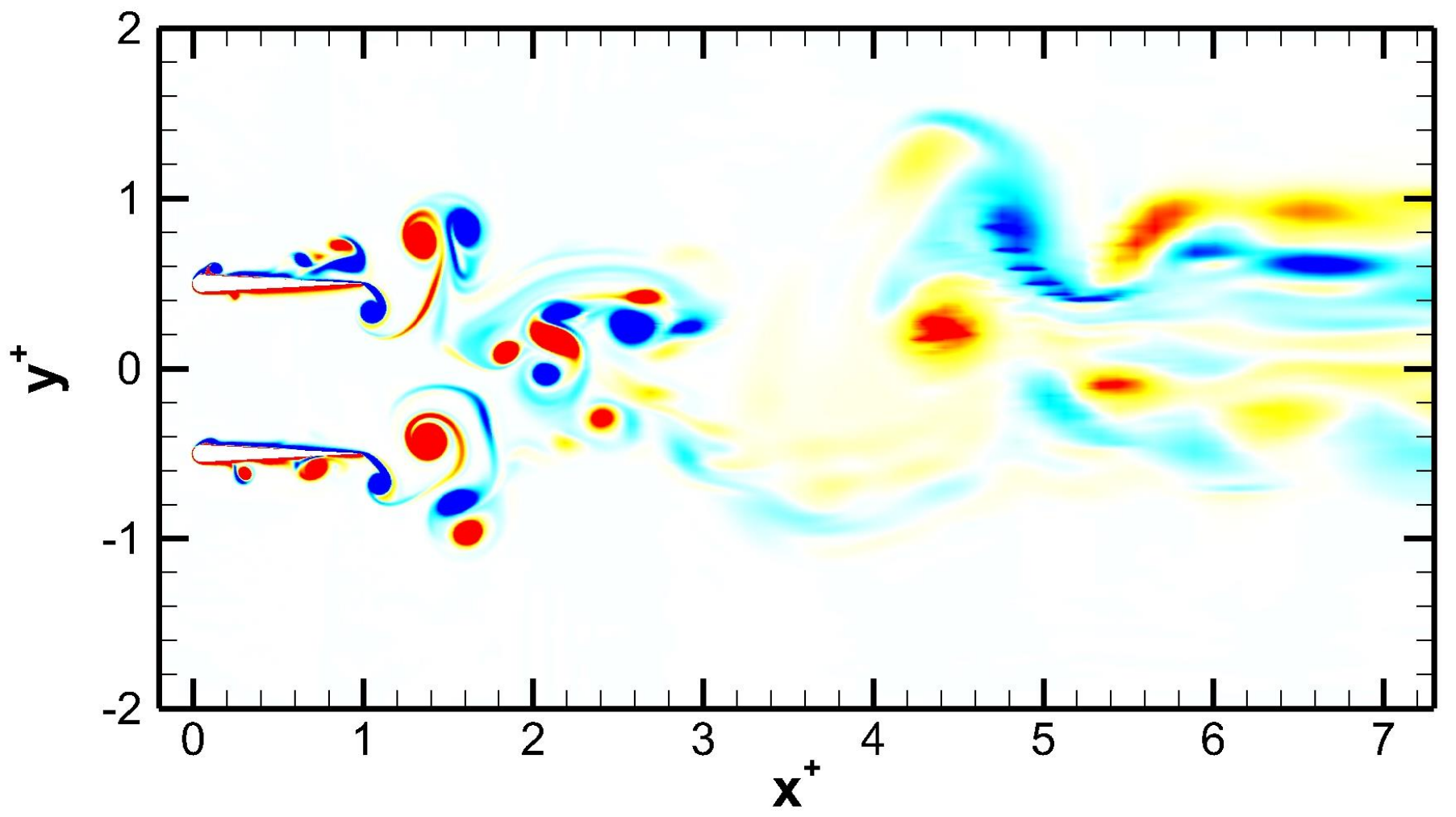}}	
\subfigure[$\quad t_3=26\tau$]{\includegraphics[width=0.47\textwidth]{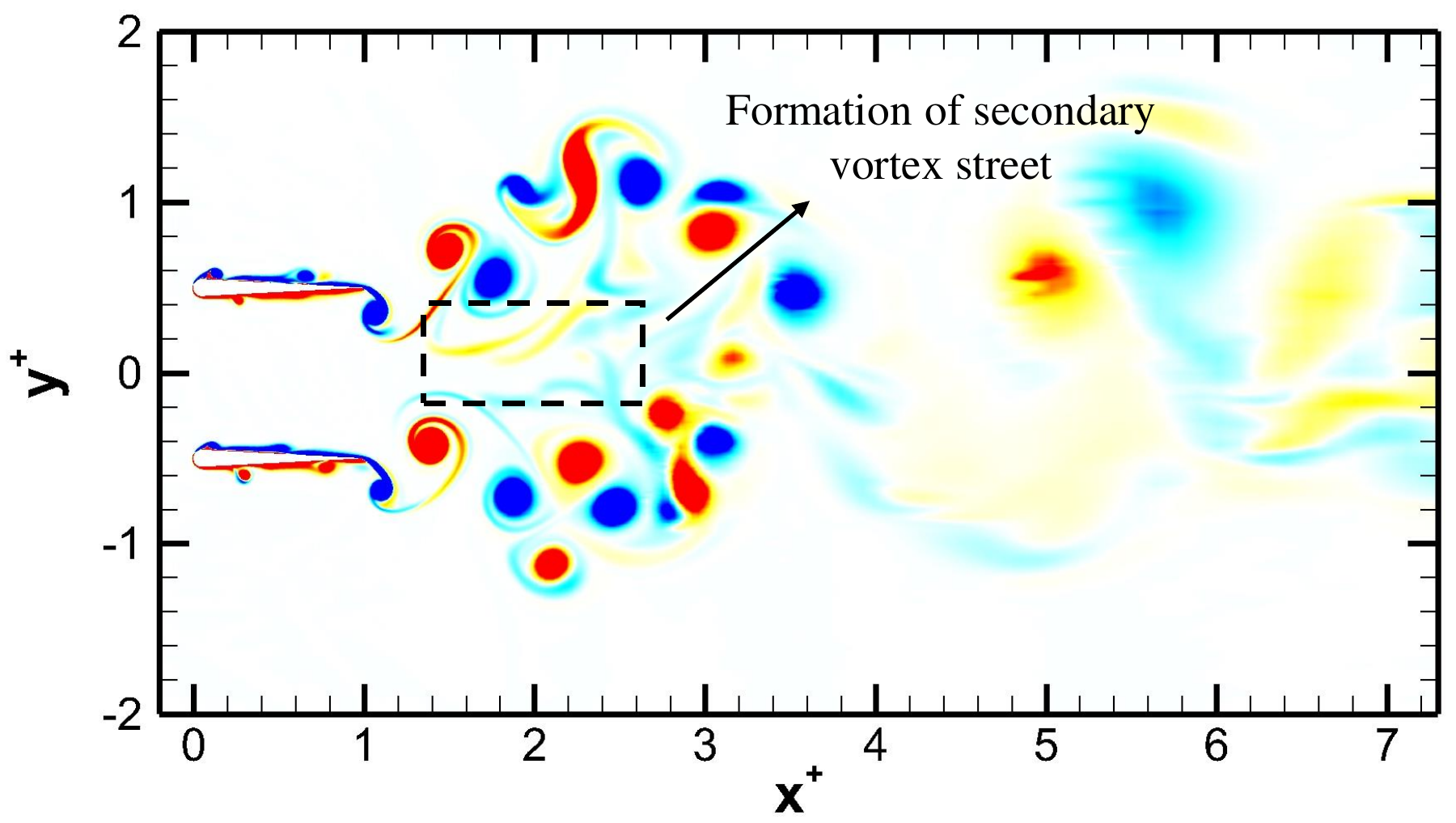}}
\subfigure[$\quad t_4=32\tau$]{\includegraphics[width=0.47\textwidth]{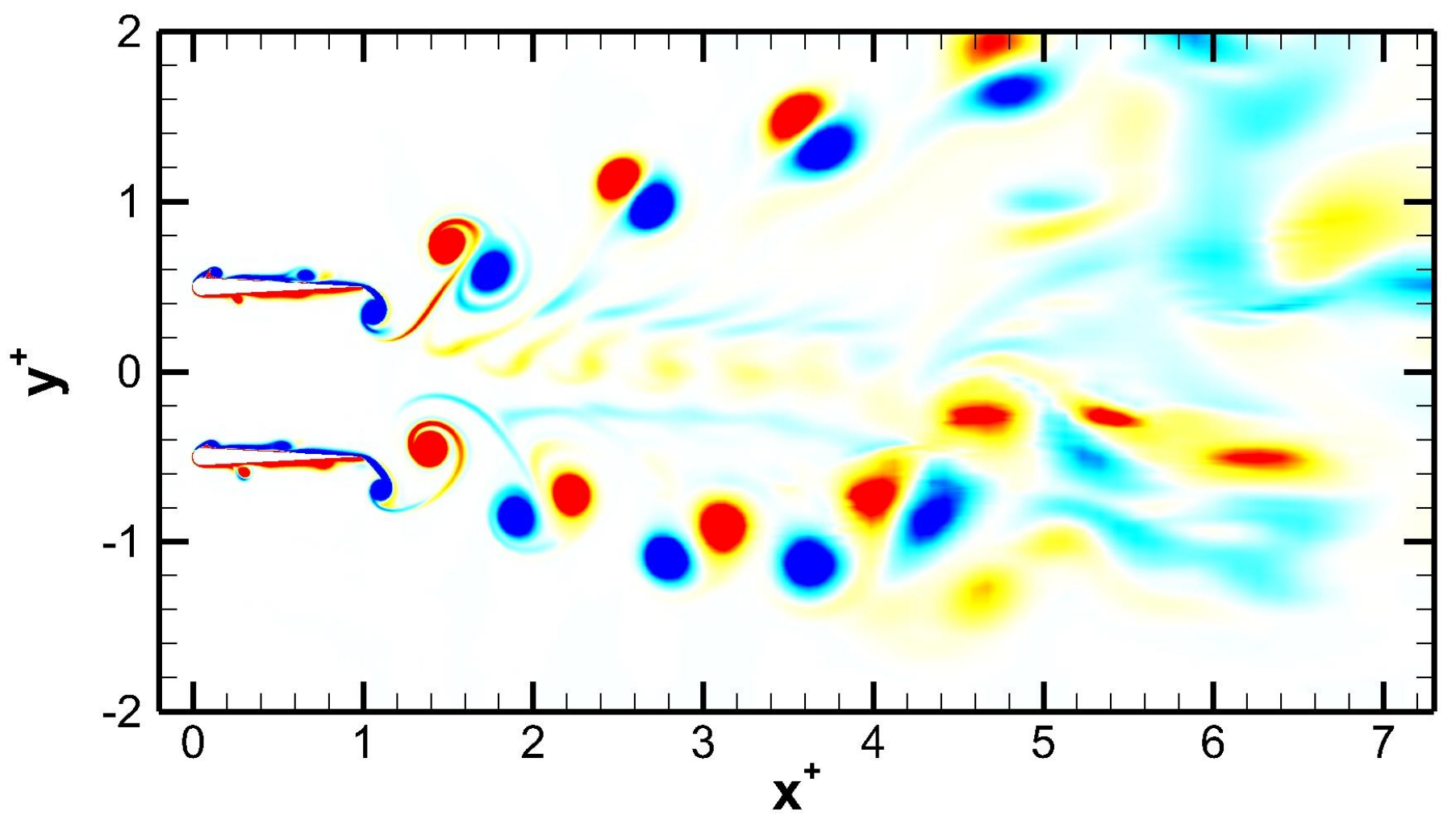}}	
\subfigure[$\quad t_0=36\tau$]{\includegraphics[width=0.47\textwidth]{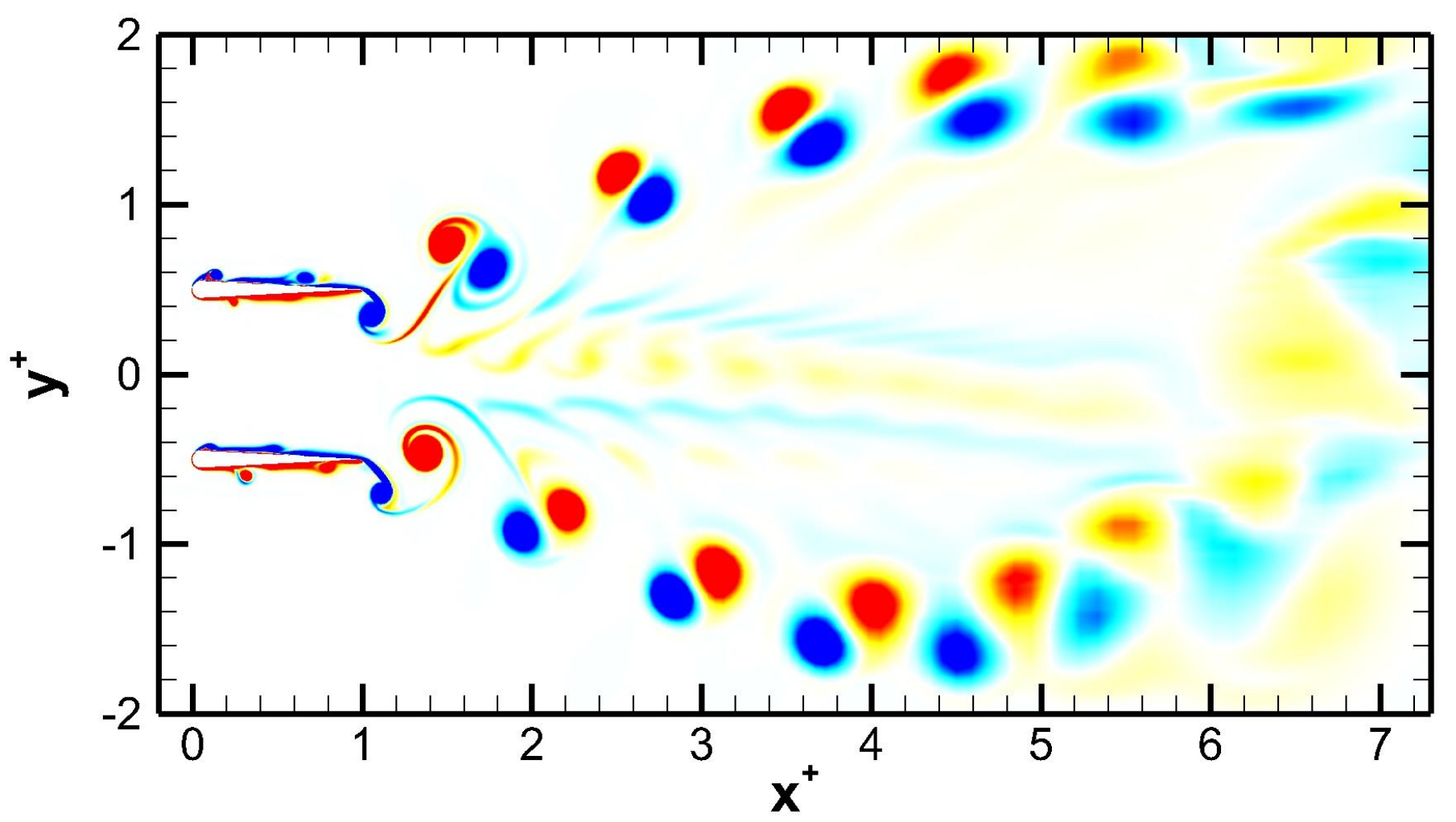}}
\hspace{7mm} 
\subfigure[$\quad t_5=40\tau$]{\includegraphics[width=0.47\textwidth]{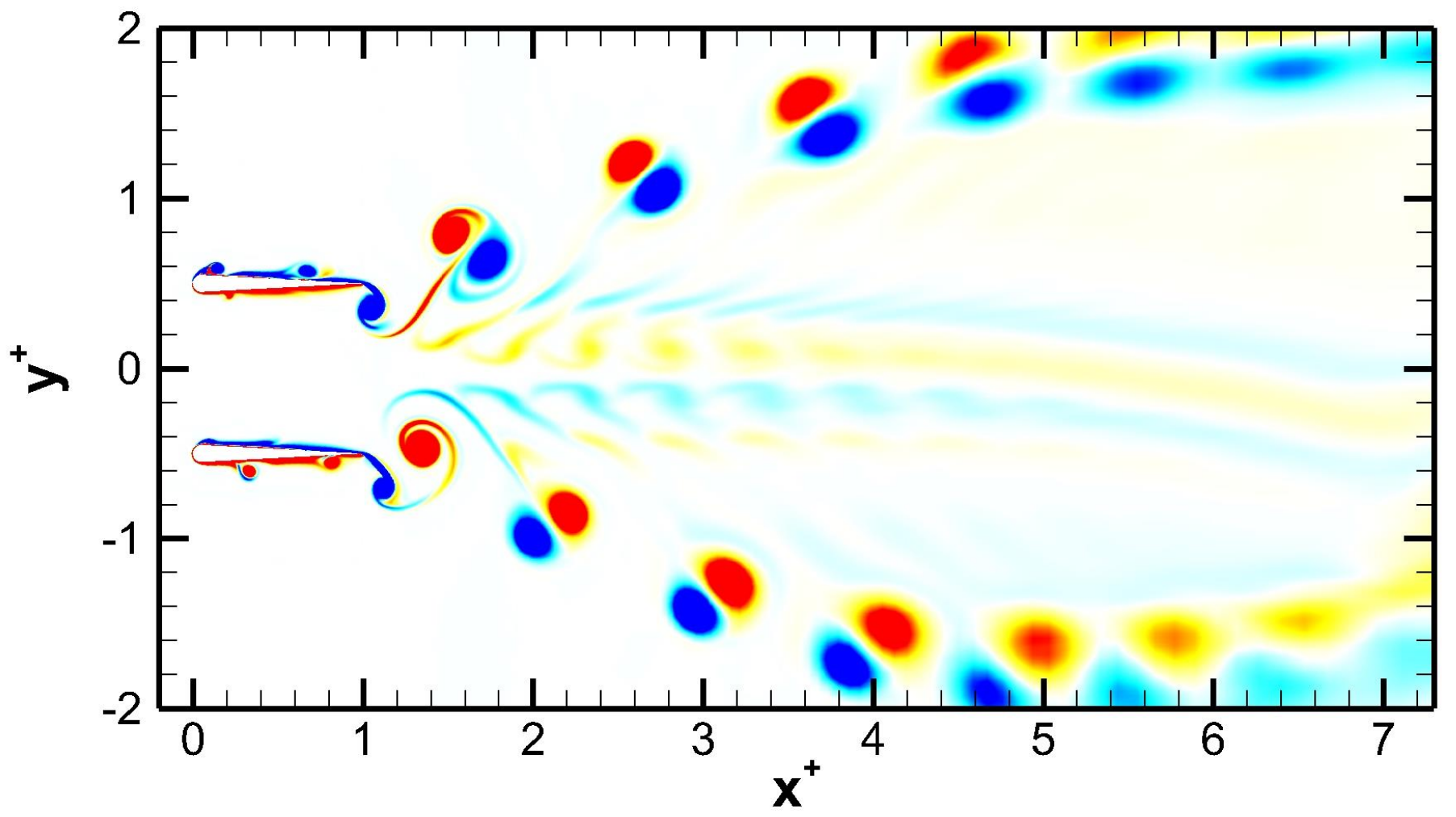}}			
\caption{Contour of instantaneous spanwise vorticity ($\omega_z^*$) over different pitching cycles for side-by-side foils following Mode 4. The contour values are the same as in Fig. \ref{fig_wake-Mode1}.}
\label{fig_wake-Mode4}
\end{figure}

The foils motion in Mode 4 only differed from that of Mode 3 on the direction of initial pitch, which was now downwards, similar to the case of Mode 2. Figure~\ref{fig_wake-Mode3}a depicts the flow dynamics for Mode 4 prior to the start of the foils in-phase pitching motion. Coherent structures are identified in the immediate vicinity of the trailing parts of both foils using dashed boundaries in Fig.~\ref{fig_wake-Mode4}a. The vortex structures shed during the $23$\textsuperscript{rd} cycle moved inwards as shown in Fig.~\ref{fig_wake-Mode4}b, but those shed in the next cycle moved outwards. During the $26$\textsuperscript{th} cycle, the secondary vortex street, represented as a dashed rectangular region in Fig.~\ref{fig_wake-Mode4}c, formed and grew downstream. Due to this phenomenon, primary vortex streets deflected from the wake center-line, that is $y^+=0$ (see Figs.~\ref{fig_wake-Mode4}d-\ref{fig_wake-Mode4}d). The only change observed after 10 cycles of in-phase pitching was that the secondary vortex street pushed the coherent structures formed between the two primary vortex streets downstream. It is important to highlight that the structure of the wake close to the foils remained the same after the $32$\textsuperscript{th} cycle, hinting at a quasi-steady behavior  (see supplementary movie 5).

\subsection{Surface Pressure Distributions and Secondary Vortex Street}

\begin{figure}
{\includegraphics[width=1.0\textwidth]{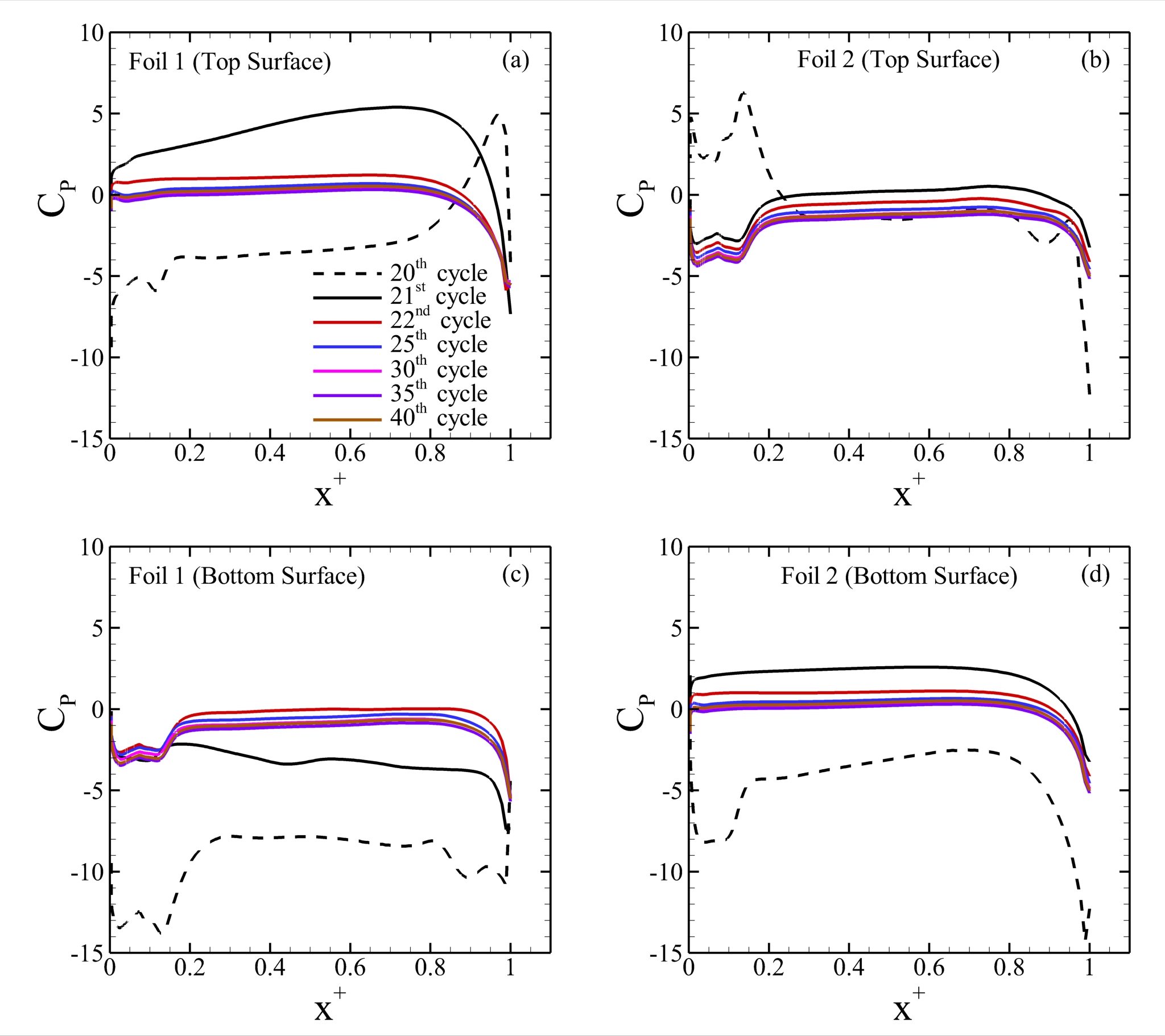}}
\caption{Cycle-averaged pressure variation on the top and bottom surfaces of Foils 1 and 2 in the case of Mode 1}
\label{fig_pressure-Mode1}
\end{figure}

Surface pressure variations were expected to dominate in alleviating side-forces. To examine this hypothesis, the phase-averaged distribution of the surface pressure coefficient ($C_p = p/0.5{\rho}{U_\infty^2}$) are shown in Fig.~\ref{fig_pressure-Mode1} for Mode 1. The profiles of pressure on upper surfaces of both foils are presented in the first row, whereas those on lower surfaces are shown in the second row. It is interesting that pressure variations on the foils approached their respective quasi-steady states in all cases faster than the wake. That it, the wake dynamics proved more complex during the same time with a quasi-steady state achieved at later stages of oscillations compared to surface pressure variations. Figure~\ref{fig_pressure-Mode1}a shows the phase-averaged pressure data on the upper surface of Foil 1. At the end of their out-of-phase oscillations in the $20$\textsuperscript{th} cycle, negative pressure was observed on $80$\% of this surface, with a large negative variation towards the foil's trailing edge, which is mostly attributed to the detachment of TEVs. After the phase was switched, surface pressure transformed to positive values for the leading $90$\% of the foil's surface, while remaining negative on the remaining part. There was a sharp decrement in pressure in the $22$\textsuperscript{nd} cycle and no significant changes were observed afterwards. A similar pattern was exhibited by pressure on the lower surface of Foil 1. Considering the side-force mathematically defined by the area between respective curves for each cycle in Figs.~\ref{fig_pressure-Mode1}a and \ref{fig_pressure-Mode1}b, it depicts mitigation of side-force right after the $22$\textsuperscript{nd} pitching cycle. 

For Foil 2, pressure was positive on the leading $20$\% part of the upper surface, whereas it became negative for $x^+ > 0.20$. For its whole lower surface, pressure remained negative. Similar to Foil 1, the pressure distribution remained the same for all pitching cycles following the $22$\textsuperscript{nd} period. The only side-force likely to be experienced by Foil 2 was due to small pressure differences on the leading $18$\% of upper and lower surfaces. Hence, quantitative and qualitative analyses demonstrated here that abrupt change of the phase angle attributes to the mitigation of side-force experienced by the system of foils, similar to solitary swimmers, due to wake deflection. Similar pressure related phenomena were also observed around the two foils for the remaining hybrid modes, which are not shown here for brevity.     

\begin{figure}
{\includegraphics[width=1.0\textwidth]{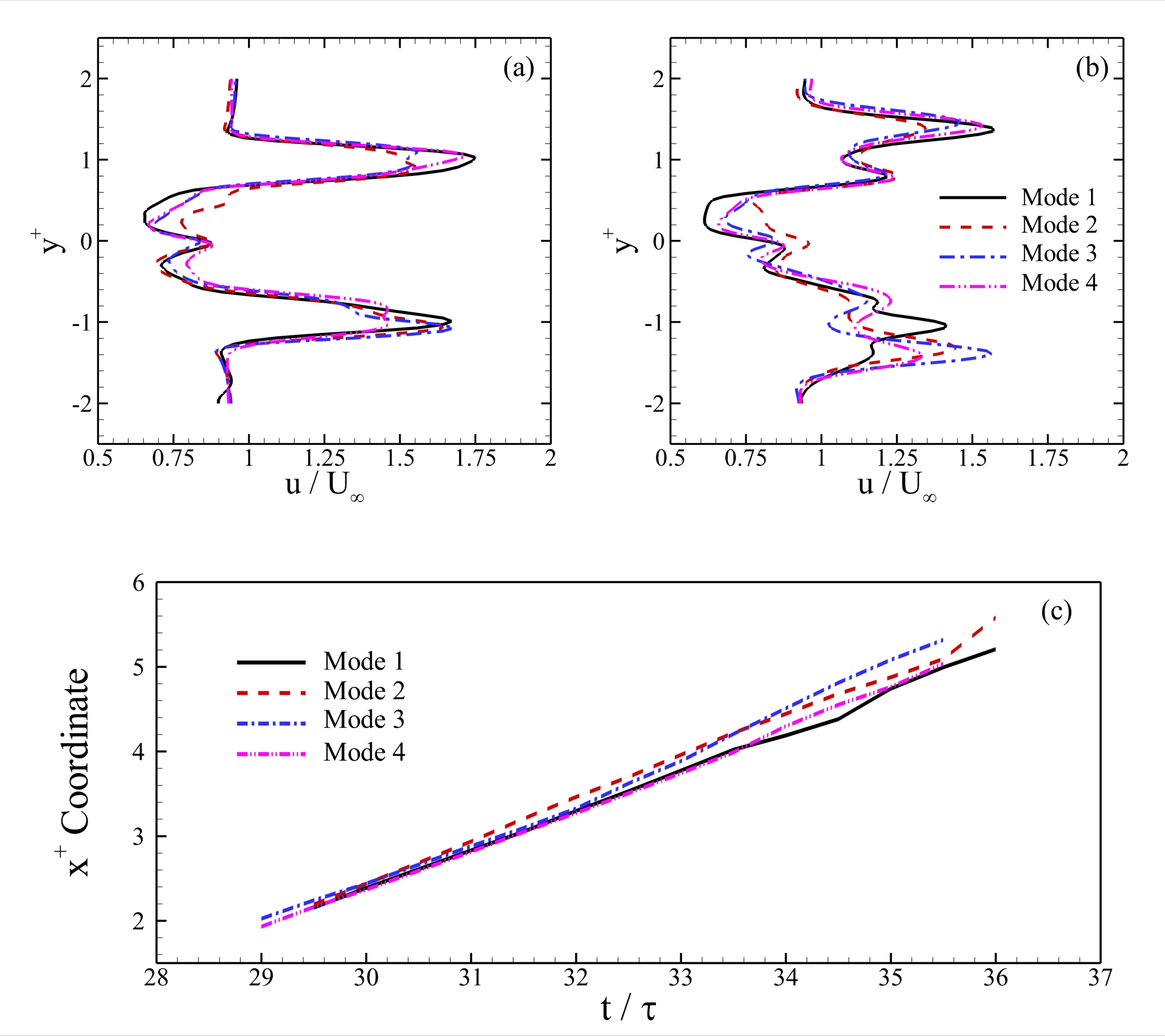}}
\caption{(a) Normalized stream-wise velocity profiles in the wake at (a) $x^+ = 2.50$ and (b) $x^+ = 3.50$, (c) $x^+$-coordinates of the leading coherent structures of the secondary vortex streets for all four hybrid modes}
\label{fig_sec_vortex_street}
\end{figure}

To examine the dynamics of the secondary vortex street formed in all hybrid modes, the time-averaged stream-wise velocity profiles are presented at $x^+ = 2.50$ and $x^+ = 3.50$ in Figs.~\ref{fig_sec_vortex_street}a and \ref{fig_sec_vortex_street}b, respectively. It is evident that the presence of the secondary vortex streets attributed to velocity deficits for all four modes. These momentum-deficit regions were responsible for the lower thrust generation by the two foils compared to that for the foils undergoing pure out-of-phase motion \citep{gungor-PRE-2020}. The upper and lower momentum-surfeit regions exist here due to the two primary vortex streets formed by both foils. The overall (total) thrust generation was also adversely affected after the switching of phase angle as manifested by the reduction in the strength of these velocity-surfeit regions compared to those presented by \cite{gungor-PRE-2020} for purely out-of-phase pitching kinematics. Figure~\ref{fig_sec_vortex_street}b presents the stream-wise velocity profiles for all hybrid modes considered in this study at $x^+ = 3.50$. It was observed that the primary vortex streets have slightly shifted their positions downstream, which was also notices in the vorticity contours of Fig.~\ref{fig_wake-Mode1}. Velocity profiles in Figs.~\ref{fig_sec_vortex_street}a and \ref{fig_sec_vortex_street}b for all four modes demonstrated a similarity in the overall wake dynamics as well as specifically for secondary vortex streets. It is important to mention that the secondary vortex street did not maintain its coherence for pure out-of-phase pitching oscillations, not shown here for brevity but may be seen in Fig.~$8$ of \cite{gungor-PRE-2020}. Its formation was observed after around $25$ oscillation cycles. However, it did not bifurcate the wake and diffused quickly providing an opportunity for the two primary vortex streets to interact with each other. More details on this wake process was discussed in \cite{gungor-PRE-2020}.

In order to understand the growth rate of the secondary vortex streets for all four hybrid modes, the coherent structures in the secondary vortex streets were visualized to track the motion of the leading vortices (positive vortex for Mode 4 and negative vortices for the remaining modes) in the wake. The abscissa coordinates of the core of these vortices are shown in Fig.~\ref{fig_sec_vortex_street}c. It was apparent that the growth rate of secondary vortex streets is almost the same for all hybrid modes and their leading vortices traversed by the same distance, although the phase was switched from $\phi = \pi$ to $0$ in different ways. The results thus far revealed an underlying flow physics in the wake of a simplified side-by-side foil system due to abrupt changes in their phase angle, which involve the formation and interaction of a secondary vortex street. This wake mechanism could be a contributing factor in fish switching their synchronization to adjust to external environmental changes or to maintain their orientation.  

\section{SUMMARY}
\label{sec_conc}

Numerical simulations of the flow over two foils pitching in side-by-side arrangement revealed unique unsteady wake modifications induced by abrupt changes in the phase angle between them at $Re=4000$ and $St=0.5$. Four different hybrid modes were considered, where the phase angle of oscillations changed from $\pi$ to $0$. This study was inspired from previously reported experimental observations for schools of red-nose fish, where individual members of these configurations changed their synchronization (phase angle) while swimming. In Mode 1 and Mode 2, foils switched their motion from pure out-of-phase pitching to in-phase motion by performing the upstroke and down-strokes first, respectively. For the other two modes, the two foils remained stationary for two oscillation periods before undergoing in-phase kinematics. The results revealed that the mid-oscillation change of phase angle resulted in the formation of a secondary vortex street between the two primary vortex streets behind individual foils. This secondary vortex street suppressed any interactions between the two primary streets in the wake due to existing induced flow by the formation and movement of vortical structures. Further study of the surface pressure variations showed that changes in the phase angle reduced the pressure difference, which attributed to alleviating side-forces for the overall dynamical system. Moreover, velocity profiles identified major changes to the wake due to the formation of, and dynamics related with, the secondary vortex streets, following the abrupt change of the foils synchronization (phase angle). It was also revealed that this new vortex street grows at the same rate for all four hybrid modes. Thus, its dynamics appeared independent from the initial dynamics associated with the foils motions, whereas there exists an ongoing debate on how such behaviors induce wake deflections differently for uninterrupted pitching foils.   

\section*{ACKNOWLEDGMENT}
This study has received support from Future Energy Systems (through Canada First Research Excellence Fund) with project number T14-Q01. Computational resources from Compute Canada and Tiger Cluster at Princeton University were used for the simulations.

\bibliography{harvard}

\end{document}